\documentclass[twocolumn,aps,floatfix,showpacs,superscriptaddress,nofootinbib]{revtex4}

\usepackage[latin9]{inputenc}
\usepackage{color}
\usepackage{verbatim}
\usepackage{amsmath}
\usepackage{amssymb}
\usepackage{graphicx}
\usepackage{esint}
\usepackage[colorlinks=true, citecolor=blue, linkcolor=blue]{hyperref}

\usepackage{xifthen}
\usepackage{booktabs}
\usepackage{ulem}

\makeatletter

\@ifundefined{textcolor}{}
{
 \definecolor{BLACK}{gray}{0}
 \definecolor{WHITE}{gray}{1}
 \definecolor{RED}{rgb}{1,0,0}
 \definecolor{GREEN}{rgb}{0,1,0}
 \definecolor{BLUE}{rgb}{0,0,1}
 \definecolor{CYAN}{cmyk}{1,0,0,0}
 \definecolor{MAGENTA}{cmyk}{0,1,0,0}
 \definecolor{YELLOW}{cmyk}{0,0,1,0}
 \definecolor{marco}{rgb}{0.7,0.,1}
}

\usepackage{dsfont}\usepackage{bm}\usepackage{bbm}

\DeclareMathOperator{\e}{e}

\newcommand{\bra}[1]{\langle#1|}
\newcommand{\ket}[1]{|#1\rangle}

\newcommand{\up}{\uparrow}
\newcommand{\down}{\downarrow}

\makeatother

\begin{document}

\title{Strongly correlated states of trapped ultracold fermions 
in deformed Landau levels}

\author{M. Burrello}
\affiliation{Max-Planck-Institut f\"ur Quantenoptik, Hans-Kopfermann-Strasse 1,
D-85748 Garching, Germany}

\author{M. Rizzi}
\affiliation{Johannes-Gutenberg-Universit\"at Mainz, Institut f\"ur Physik, 
Staudingerweg 7, D-55099 Mainz, Germany}

\author{M. Roncaglia}
\affiliation{INRIM, Strada delle Cacce 91, 10135 Torino, Italy}

\author{A. Trombettoni}
\affiliation{CNR-IOM DEMOCRITOS Simulation Center, Via Bonomea 265, I-34136
Trieste, Italy}
\affiliation{SISSA and INFN, Sezione di Trieste, Via Bonomea 265, I-34136 
Trieste, Italy}

\begin{abstract}

We analyze the strongly correlated regime of a two-component trapped ultracold fermionic gas
in a synthetic non-Abelian $U(2)$ gauge potential,
that consists of both a magnetic field and a homogeneous spin-orbit coupling.
This gauge potential deforms the Landau levels (LLs) with respect to the Abelian case
and exchanges their ordering as a function of the spin-orbit coupling. 
In view of experimental realizations, we show that a harmonic potential
combined with a Zeeman term, gives rise to an angular momentum term,
which can be used to test the stability of the correlated states obtained through interactions. 
We derive the Haldane pseudopotentials (HPs) describing the interspecies contact interaction
within a lowest LL approximation.
Unlike ordinary fractional quantum Hall systems and 
ultracold bosons with short-range interactions in the same gauge potential,
the HPs for sufficiently strong non-Abelian fields show an unconventional 
non-monotonic behaviour in the relative angular momentum. 
Exploiting this property, we study the occurrence of new incompressible ground states 
as a function of the total angular momentum. 
In the first deformed Landau level (DLL) we obtain Laughlin and Jain states.
Instead, in the second DLL three classes of stabilized states appear: Laughlin states, 
a series of intermediate strongly correlated states and finally vortices of the integer quantum Hall state.
Remarkably, in the intermediate regime, the non-monotonic HPs of the second DLL induce two-particle correlations which are 
reminiscent of paired states such as the Haffnian state. 
Via exact diagonalization in the disk geometry,
we compute experimentally relevant observables such as density profiles and correlations,
and we study the entanglement spectra as a further tool to characterize the 
obtained strongly correlated states.

\end{abstract}

\pacs{73.43.-f, 67.85.Lm}

\maketitle

\section{Introduction}

Ultracold atoms offer a versatile and controllable environment
to tailor a huge variety of many-body quantum systems~\cite{bloch08}.
Motivated by the remarkable success in the experimental implementation
of several solid state systems with such platforms~\cite{lewensteinbook},
it seems fruitful to push further the implementation and the study of paradigmatic models, 
whose realization in solid state physics is limited to certain ranges of the physical parameters.

One such case is the realization of fractional quantum Hall (FQH) states,
whose physics in semiconductor setups is governed by the Coulomb repulsion between electrons.
Due to its fundamental character, this interaction is very hard to tune, 
therefore several interesting many-body ground states,
theoretically predicted for particles in Landau levels (LLs),
cannot be unambiguously achieved and/or tested.
In this respect, the main advantage in simulating the strongly correlated quantum Hall physics 
on cold atomic platforms would reside in the high and precise tunability of two-body interactions.
A wealth of physical regimes that have been hitherto elusive in the corresponding semiconducting systems
might become attainable, such as the design and control of fractional quantum Hall states~\cite{cooper01}.
A further important application concerns optical lattice systems with  
non-Abelian gauge potentials, where an anomalous quantum Hall effect~\cite{goldman09a,goldman09b} may appear.

In real experiments, reaching the quantum Hall regime in ultracold gases is still an open challenge. 
In the last few years several milestone progresses towards this achievement have been accomplished~\cite{goldman13}:
in particular it has been experimentally proved that careful designs of Raman couplings
allow to simulate the effect of artificial electric~\cite{spielman09b} and magnetic~\cite{spielman09a,zu11} fields
by providing the single-particle wave functions with non-trivial Berry phases;
moreover, also the realization of spin-orbit coupled gases, and thus non-Abelian gauge potentials,
has been achieved for both bosons~\cite{spielman11} and fermions~\cite{wang12,zw12}.
Very recently, different techniques have been adopted to create a synthetic magnetic field in 
optical lattices through either laser dressing~\cite{spielman12,bloch11,bloch13,ketterle13,rec14} 
or shaking of the lattice~\cite{struck12,esslinger14}.

Motivated by these achievements, it is certainly interesting to discuss the feasibility of realizing 
strongly correlated states using a non-Abelian $U(2)$ gauge potential 
which can be tailored to implement the simultaneous presence of a magnetic field and a 
homogeneous spin-orbit coupling.
A generic non-Abelian term wipes away the LL structure~\cite{santos07,clark08}.
It was however shown~\cite{burrello10,burrello11} that this is not the case of the particular $U(2)$ potential under consideration 
in the present paper.
The LL structure is indeed preserved and simply undergoes a deformation,
that splits the degeneracy among internal states and leads to a sequence of different LLs as lowest energy states
as a function of the spin-orbit coupling. 
The effective interaction between atoms in these deformed Landau levels (DLLs) 
can be described in terms of Haldane pseudopotentials (HPs) \cite{haldane}
(i.e., one coefficient per each relative angular momentum between particles),
as proposed in diverse physical systems displaying an effective LL 
structure~\cite{jain,chakraborty11,papic11,dasilva11,simon12}.
In the case of ultracold gases subjected to the considered $U(2)$ gauge potential,
it has been shown that even a bare contact interaction between the atoms yields two-particle
HPs in the DLLs that dramatically vary with the spin-orbit coupling
and introduce effective finite-range interactions~\cite{burrellothesis,cooper12}.

The effect of the HP form in such a DLL framework has been 
quite extensively investigated for bosonic particles,
both in a lowest DLL approximation~\cite{pachos} as well as
close to the crossing points of the DLL structure~\cite{grass12a,ueda,grass12b},
where it can be compared to the usual two-component 
gas (i.e., without non-Abelian terms)~\cite{ueda13,wu13,regnault13,grass13a}.
In particular, exact diagonalization studies in a torus geometry~\cite{grass12a,grass12b} have shown
that the $U(2)$ non-Abelian gauge potential favours the appearance of 
bosonic versions of the Read-Rezayi~\cite{readrezayi,cappelli01} 
and non-Abelian spin-singlet~\cite{ardonne99,ardonne01} FQH states,
respectively in the non-degenerate and degenerate regime of DLLs.
For ultracold bosons the dynamics and the detection of quasiholes were also investigated~\cite{grass12c,grass13}
and an analysis of the edge states was reported in~\cite{grass13a}. 

In this paper we consider instead a {\it fermionic} gas subject to the same synthetic $U(2)$ gauge potential
and to contact interactions among its two bare spin components.
Specifically, we focus on the non-degenerate DLL regime, where the system can be mapped
into an effective single-component gas, whose character depends on the spin-orbit coupling intensity.
We show that contact interactions are mapped in non-trivial finite-range interactions allowing, for example, to give rise to different Laughlin states without recurring to long-range interactions as customarily done for single-component fermionic gases \cite{grass11,lewenstein13}.

Our main goal is twofold:
on one side, we aim at characterizing the specific properties of interacting fermions not 
reproducible in a Bose gas subject to the same gauge potential.
The most striking of such features is that from simple standard contact interactions it is possible to get HPs that 
do not decay monotonically as a function of the relative angular momentum.
On the other side, we discuss how to take advantage of the presence of a harmonic confinement
to generate a canonical angular momentum term, which stabilizes certain strongly correlated, incompressible, states.

To this purpose we adopt in our exact diagonalization scheme the disk geometry, 
which is more relevant for the analysis of realistic implementations. 
Although the effect of harmonic trapping on the quantum Hall regime of ultracold atomic gases
has been already analyzed for several bosonic settings~\cite{cazalilla05,cooper08,fetter,grass12,pino13},
it remains overall unexplored for ultracold fermions.

By varying the spin-orbit coupling, we study both sides 
of the point at which the two lowest DLLs cross (see Fig. \ref{fig:spectrum}) and alternate 
as single-particle ground states in such a way that the different 
behaviour of the HPs yields to rather different many-body ground states. 
In particular we focus on two suitably chosen values of the strength 
of the non-Abelian potential such that the smaller (larger) value 
presents the first (second) DLL as single-particle ground state. 
The first DLL is characterized by a single HP: therefore, the stabilised many-body solutions are
the $1/3$ Laughlin state in the limit of small trapping 
and the usual Jain hierarchical states at stronger confinement. 
Within the second DLL, instead, two {\it non-monotonic} HPs are present
and three classes of stabilized states are obtained: 
$1/5$ Laughlin states, a series of intermediate strongly correlated states 
reminiscent of the Haffnian~\cite{wen94} and the W\'ojs, Yi and Quinn states~\cite{wojs04},
and finally vortices of the integer quantum Hall droplet.
For all these states we have also calculated observables, 
such as density profiles and correlations, which are directly measurable in experiments. 
Furthermore, we present a study of the particle entanglement spectrum in the 
obtained strongly correlated states as a further tool for their characterization.

The paper is organized as follows.
In Sections \ref{sp} and \ref{HPs} we provide a detailed construction of the Hamiltonian
with a non-Abelian gauge potential in its single- and two-body (pseudo-potential) part, respectively.
Furthermore, in Section \ref{sp} remarks on the experimental feasibility are provided,
while in Section \ref{HPs} we sketch the Hamiltonian form employed for numerical purposes.
In Section \ref{below} we consider the first DLL
and analyse the stability window for FQH states belonging to the Jain hierarchy.
In Section \ref{above} we focus instead on the second DLL, 
with its two non-monotonic HPs playing a central role to induce non-standard stable ground-states:
their appearance is examined by signatures in the (spin-resolved) density profiles and two-body correlation functions.
In Section \ref{ent_sp} we investigate the particle entanglement spectrum of such states,
and propose some conjectures about their nature.
A summary and some outlook are then presented in Section \ref{concl}.
Further details about the construction of pseudopotentials and the numerical
algorithm are reported in the Appendices \ref{app_dec} and \ref{app_num} respectively. 

\section{Single-particle Hamiltonian}\label{sp}

We study a trapped gas of ultracold fermions in a two-dimensional geometry,
obtained by a strong confinement along the $z$ direction.
The gas is also subjected to an in-plane harmonic trapping with frequency $\omega$.
The atoms have an internal two-component degree of freedom,
which plays the role of an effective spin $1/2$ and may correspond, for instance,
to hyperfine levels or to degenerate dark states coupled by a suitable optical setup.
These spin degrees of freedom are coupled to the orbital dynamics through
the introduction of an external non-Abelian gauge potential $\vec{\cal A}$,
in the form of $2\times2$ matrix-valued components.
We also consider a supplementary spin-dependent
potential term $V_s$ that will be specified in the following.
The single particle $2D$ Hamiltonian reads then
\begin{equation}\label{ham2_in}
 H=\left(p_{x} \mathbb{I} +{\cal A}_{x}\right)^{2}+\left(p_{y} \mathbb{I} +{\cal A}_{y}\right)^{2} + 
\frac{\omega^2 r^2}{4} \mathbb{I} + V_s(x,y),  
\end{equation}
where $\vec{p}$ is the particle momentum, $\mathbb{I}$ is the $2\times2$ identity matrix,
and we set the atomic mass to $m=1/2$ (and as well $\hbar=1$).

In this work we consider the $U(2)$ potential arising from the interplay between
a spin-independent $U(1)$ synthetic magnetic field with intensity ${\cal B}$ and
a $SU(2)$ spin-orbit coupling of constant strength $q$. In the symmetric gauge the vector potential then reads:
\begin{equation}\label{eq:A}
 \vec{{\cal A}}=\left(	-\frac{y {\cal B}}{2}\mathbb{I}+q\sigma_{x};	\frac{x{\cal B}}{2} \mathbb{I}+q\sigma_{y},;	0	\right).
\end{equation}
At variance with the purely Abelian case, the spatial components of the $F^{\mu \nu}$ tensor 
do not define a physical, gauge-invariant, (pseudo-)vector field.
Indeed the same $\vec{F} =  \vec{\nabla}\times\vec{{\cal A}}+
i \vec{{\cal A}}\times\vec{{\cal A}} = \left( {\cal B} \, \mathbb{I} - 2 q^2 \sigma_z \right) \hat{z}$
could be as well obtained through a very different setup, namely two uncoupled spin species subjected to 
different magnetic fields.
Nevertheless, this latter situation would instead correspond to an Abelian $U(1) \times U(1)$
gauge potential giving rise to a trivial coupling between the spins~\cite{brown79,estienne11}.

In presence of a weak harmonic confinement $\omega~\ll~{\cal B}$ (see Sec.~\ref{ssec:exp} for details),
we can conveniently recast Eqs.~(\ref{ham2_in}-\ref{eq:A}) in terms of an effective magnetic field $B$
(vector field $\vec{A}$, respectively) and its shift $\Delta$ from the bare synthesised one ${\cal B}$:
\begin{equation}
	B \equiv\sqrt{{\cal B}^2+\omega^2} \,, \qquad 
	\Delta \equiv B-{\cal B} \sim \omega^2/ 2 B \, .
\end{equation}
Two further terms are originated by the confinement:
first, an angular momentum contribution of the kind $- \Delta (x p_y - y p_x)$, 
i.e. proportional to $L_z$;
second, a Zeeman term $+ q \Delta \left(y\sigma_x-x\sigma_y\right)$, which varies linearly in space.
The latter can be compensated by a proper choice of the supplementary potential (see also \cite{xu10,zhou11}):
\begin{equation}\label{additional}
 V_s(x,y)=-q\Delta\left(y\sigma_x-x\sigma_y\right) \, ,
\end{equation}
attainable by a {\it true} magnetic field gradient~\cite{ketterle13,bloch13} 
coupling to the internal degrees of freedom.
The final form of the single-particle Hamiltonian, subject of investigation in the rest of our paper, reads:
\begin{equation} \label{ham2}
 H= \left(p_x -\frac{yB}{2} +q \sigma_x\right)^2 + \left(p_y+\frac{xB}{2} +q\sigma_y \right)^2 -L_z \Delta .
\end{equation}

\subsection{The deformed Landau Level structure} \label{ssec:DLL}

The single-particle Hamiltonian \eqref{ham2} was studied in \cite{burrello10,burrello11}:
we recall here the most relevant results. 
As a first step, let us define the canonical wave functions of the magnetic ($q=0$) LLs
in the symmetric gauge: 
\begin{equation} \label{eq:LL}
\psi_{n,m}=i^{n}\sqrt{\frac{2^{n}n!}{2\pi2^{m}m!}}z^{m-n}L_{n}^{m-n}\left(|z|^{2}/2\right)\e^{-B|z|^{2}/4},
\end{equation}
where $z=x-iy$ is the complex coordinate of the atom in the $2D$ plane
and $L_{n}^{m-n}$ is a generalized Laguerre polynomial.
The integer $n$ is the LL index corresponding to the ladder operators
$d=(B z + 4 \partial_{\bar{z}})/\sqrt{8B}$ and $d^{\dag} = (B \bar{z} - 4 \partial_{z})/\sqrt{8B}$,
whereas $m$ labels states with different angular momentum $L_{z}=n-m$
within the same LL $n$. 

The non-Abelian term of~\eqref{ham2} couples the states
$\left\{ \psi_{n-1,m} \ket{\uparrow}, \, \psi_{n,m} \ket{\downarrow} \right\}$ through a Jaynes-Cummings term.
Each block can be recast into:
\begin{equation}
H_{n,m} = \tilde{\varepsilon}_{n,m} \, \mathbb{I}
		+D_{n} 	\begin{pmatrix}
					- \cos\varphi_{n} & \sin\varphi_{n}\\
					\sin\varphi_{n} & \cos\varphi_{n}
				\end{pmatrix} \, ,
\end{equation}
where we adopted the convenient parametrisation:
\begin{eqnarray} 
\label{eq:eps}
  \tilde{\varepsilon}_{n,m} & = & 2q^{2} + \Delta/2 + 2 (B - \Delta/2) n + m \Delta  \\
\label{eq:Dn}
 D_{n} & = & \sqrt{\left(B-\Delta/2\right)^{2}+8q^{2}Bn}\,,\\
\label{eq:phin}
 \cos\varphi_{n} & = & \frac{B-\Delta/2}{D_{n}} \,,\quad 
 \sin\varphi_{n} = \frac{2q\sqrt{2Bn}}{D_{n}} \, .
\end{eqnarray}
Consequently, we can describe the eigensystem of the $2~\times~2$ block-diagonal Hamiltonian
in terms of
\begin{eqnarray}\label{chidelta}
\label{eq:chi}
\varepsilon_{n,m} =  \tilde{\varepsilon}_{n,m} - D_{n} \, ,
 & \ & \chi_{n,m}  = \left( + \cos \frac{\varphi_{n}}{2}; -\sin \frac{\varphi_{n}}{2} \right) \, ; \qquad \\
\label{eq:chi1}
 \varepsilon_{n,m}^\prime =  \tilde{\varepsilon}_{n,m} + D_{n} \, ,
 & \ & \chi_{n,m}^\prime  = \left( + \sin \frac{\varphi_{n}}{2}; + \cos \frac{\varphi_{n}}{2} \right)  \, .  \qquad
\end{eqnarray}
We will adopt the short-hand notation $\varepsilon_n\equiv \varepsilon_{n,0}$.

\begin{figure}
\includegraphics[width=1\columnwidth]{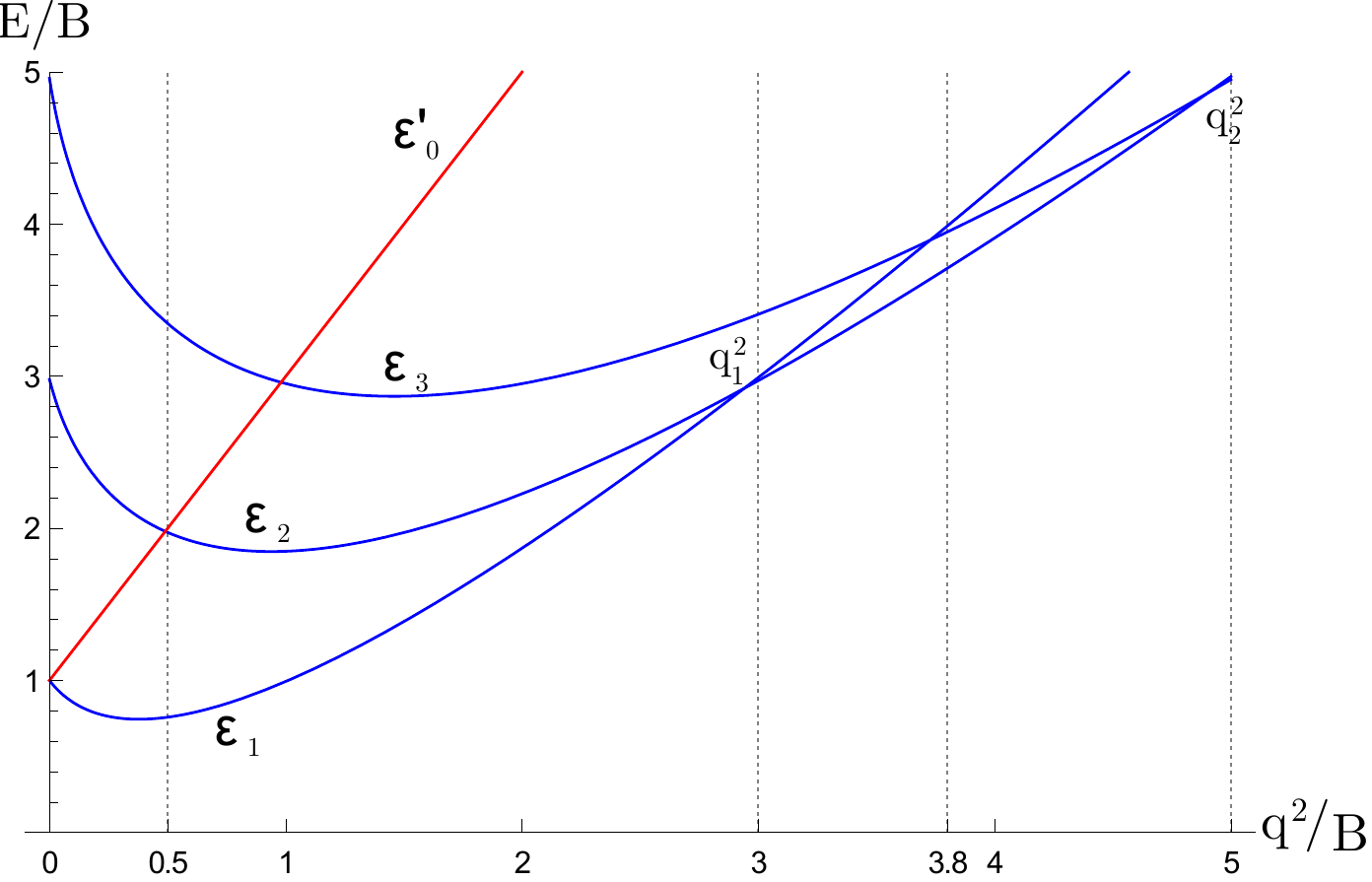} 
\caption{The energy levels 
$\varepsilon_{1},\varepsilon_{2},\varepsilon_{3}$ (blue) 
and $\varepsilon_{0}^{\prime}$ (red) are shown as a function of $q^2/B$ for 
$\Delta/B=0.02$ and $m=0$ (the energy levels $\varepsilon'_1$ and $\varepsilon'_2$ are not shown). 
The crossing points $q^2_1$ and $q^2_2$ 
are slightly displaced from their values $q^2/B=3,5$ at $\Delta=0$. 
The numerical analysis in Sections \ref{below} and  \ref{above} 
will be performed at the values $q^2/B=0.5$ and $q^2/B=3.8$ 
which are close to the maxima of the gaps between the ground state 
energy and the first excited DLL.}
\label{fig:spectrum}
\end{figure}

An unpaired set of states $\chi^\prime_{0,m} = \psi_{0,m} \ket{\down}$ is present in addition, 
but its energy $\varepsilon^\prime_{0,m} = B + 2 q^2 + m \Delta$ is higher than the other $\chi$ states,
therefore it is not relevant for our analysis.
The sets $\chi$ and $\chi'$ define the DLLs, still labeled by $n$:
their energies present several crossing points as $q$ is increased~\cite{rashba84,burrello10,burrello11},
therefore giving rise to an interchange in the role of the lowest one.
The degeneracy between the lowest energy states of the sets $\chi_{n}$ and $\chi_{n+1}$ happens at
\begin{equation} \label{qcross}
	q_{n}^{2}(B,\Delta)=\frac{\left(2n+1\right) \left(B-\frac{\Delta}{2}\right)^{2}}{B} \ \ \forall n \geq 1 \quad (q_0 = 0),
\end{equation}
only slightly displaced from the $\Delta=0$ odd-integer values $q_{n}^{2} (B,0) = (2n+1) B$,
for our regime of interest $\Delta \ll B$.
Thus the wavefunctions $\chi_{n,m}$ of Eq.~\eqref{eq:chi} describe the lowest energy states
for $q_{n-1}^{2}<q^{2}<q_{n+1}^{2}$.
As shown in Fig.~\ref{fig:spectrum}, the energy gaps between the lowest DLL
and the higher ones are
\begin{equation}\label{eq:DLLgap}
\delta_n^\pm \equiv  \varepsilon_{n \pm 1,0} - \varepsilon_{n,0} = \pm 2 (B - \Delta/2) + D_{n \pm 1} - D_n \ .
\end{equation}
For reasons that will be discussed in detail later, we are interested 
in working in the vicinity of the maxima of these gaps, which 
in this case occur in correspondence of the crossings between $\chi^\prime_{0}$ and $\chi_{2}$ for $q^{2}<q_{1}^{2}$
and $\chi_{1}$ and $\chi_{3}$ for $q^{2}>q_{1}^{2}$.
For $\Delta=0$, these crossings are at 
$q^{2}=B/2$ and $q^{2}=\left(4+\sqrt{13}\right)/2 B \approx 3.8 B$~\cite{burrello11}, 
which will then constitute the optimal points for the numerical analysis
of many-body states in Secs.~\ref{below} and \ref{above}, respectively. 

These maximal gaps, for $n>1$, can be approximated by $B/(2n)$: this allows us to give a first estimate of the validity of the Landau level description as a function of the trapping. By considering a system with $N$ atoms at filling $\nu$, the highest occupied orbital corresponds approximately to $m_{\rm max}=N/\nu$; thus, from Eq. \eqref{eq:eps}, we find that the condition $B/(2n) \approx B^2/q^2 \gg  N \Delta/\nu $ must be fulfilled in order to avoid that excited DLLs become occupied also far from the degeneracy points. Let us consider, for instance, the typical set of value that we will adopt to examine the second DLL ($n=2$) in Sec. \ref{above}: for $q^2=3.8 B$, $N=8$ and $\nu=1/5$, the previous condition is verified for $\Delta \ll B/160$ leading to $\omega \ll \mathcal{B}/9$.

\subsection{Comments about possible experimental realizations} \label{ssec:exp}

The Hamiltonian~\eqref{ham2} results from the interplay of three main physical elements:
an artificial magnetic field, a spin-orbit coupling and a harmonic trapping potential.
A detailed analysis of the physical realization of this setup is beyond
the scope of this work and the reader may find a comprehensive description
of the implementations of artificial magnetic fields and spin-orbit couplings,
e.g., in the recent review \cite{goldman13}.
Anyway, we present here a brief overview of some experimental building blocks
that might be combined to realize the non-Abelian $U(2)$ potential,
and, based on them, we also provide some estimates of typical Hamiltonian parameters.

One of the oldest strategies to engineer synthetic magnetic fields for neutral atoms
relies on the formal equivalence between the Coriolis force in
a rotating system and the Lorentz force in a uniform magnetic field.
The magnetic field is then obtained as the interplay of the centrifugal term and the harmonic potential%
~\cite{cooper08,fetter}.
Due to unfavourable scalings of gaps with the number of particles  
and to dissipation problems arising when going to extreme rotation velocities,
it seems difficult to reach the quantum Hall regime.
		
To overcome this drawback, one can either engineer dynamical systems,
based on a ring geometry to achieve large angular momenta~\cite{roncaglia11},
or resort to optical couplings following, for instance, the approach in~\cite{spielman09c}. 
Another possibility is offered by the use of traps with a small number 
of particles which have been experimentally investigated in~\cite{chu}.
	
A more general strategy to create artificial gauge potentials relies on tailoring laser-fields
in order to get a space-dependent basis of dressed states.
The atoms moving so slowly to stay adiabatically on the ground manifold
get affected by a non-trivial Berry phase (or matrix) that resembles the desired gauge 
one~\cite{dalibard11,goldman13}.
Most of the recent experimental results on spin-orbit coupling
have been indeed obtained through a pair of Raman lasers
transferring a recoil momentum to the atoms which depends on their internal 
state~\cite{spielman11,wang12,zw12,galitski2013}.
This is however limited to the creation of a spin-orbit coupling
along a single direction (equal superposition of Rashba and Dresselhaus).
Nonetheless, more sophisticated schemes have been put forward
to create a full two-dimensional SU(2) potential~\cite{campbell2012} 
or even a three dimensional spin-orbit coupling~\cite{anderson2012} 
(notice that realizing a 3D anisotropic spin-orbit coupling 
$\vec{A}\propto(q \sigma_x, q\sigma_y,q_z \sigma_z)$  one could obtain the non-Abelian $SU(2)$ term for $q_z \to 0$).
	
The simultaneous presence of a spin-independent magnetic field
with a non-Abelian spin-orbit term, as needed here for the gauge potential of Eq.~\ref{eq:A},
has been shown to be achievable in rotating systems~\cite{burrello11,galitski11}
based on the so-called tripod atom-lasers schemes~\cite{ruse05}.
Moreover, the Zeeman contribution $V_s$ in Eq. (\ref{additional}) can be obtained by adding a 
physical magnetic field that varies linearly in space, as already done in the experimental setups in 
Refs. \cite{ketterle13,bloch13}, for instance. 
Let us stress that also non-optical techniques can be used to generate effective spin-orbit terms:
for example, the Zeeman splitting of a pulsed and inhomogeneous magnetic field 
has been successfully used to produce a Rashba coupling~\cite{anderson2013,ueda13b}.

By combining the typical values of the above mentioned setups,
we are now able to provide some rough estimates of the values 
achievable by the parameters of the single-particle Hamiltonian~(\ref{ham2_in}-\ref{ham2}). 
The value of ${\cal B}$ in recent lattice experiments~\cite{rec14} is indeed such that 
${\cal B} \ell^2 \simeq \hbar$, giving ${\cal B} \simeq 4  \cdot 10^{-19} g/s$ 
for a typical alkali lattice spacing $\ell \simeq 0.5 \mu m$.
On one hand, when compared to shallow trapping frequencies between $10$ and $100$ Hz
and the mass of $K$ atoms, we obtain that $m\omega / {\cal B} \simeq 0.01 \, \div \, 0.1$
and consequently $\Delta / {\cal B} \simeq 0.005 \div 0.05$.
On the other hand, spin-orbit momenta of the order $\hbar / q \simeq 1\mu m$
have already been demonstrated~\cite{spielman11},
allowing to safely speculate on values $q^2 / \hbar {\cal B} \simeq 0.1 \div 5$,
as needed to address the first and second DLL in the present work.

\section{Haldane pseudopotentials}\label{HPs}

The scattering properties of ultracold atoms can be efficiently modelled by means of
a contact repulsion with a typical energy scale $v$ ($s-$wave approximation).
In the following we show that
the above computed deformation of Landau Levels
effectively leads to longer-ranged interactions between particles sitting on them,
and therefore to non-trivial strongly correlated phases.
In order to see that, let us revise (and specify to our setup) the two main assumptions
that are commonly used to study interacting gases within a LL picture:
namely, the lowest LL approximation and the formalism of HPs~\cite{haldane}.
Such a path has been demonstrated successful in a wealth of physical systems,
ranging from usual quantum Hall setups (see, e.g.,~\cite{jain} and references therein)
to graphene~\cite{chakraborty11,papic11,simon12} and topological insulators~\cite{dasilva11}.

The two-species mixture can be treated as a single-component gas living on the lowest DLL $\chi_n$,
provided the occupation of higher DLLs can be neglected.
This translates into two conditions for the DLL gap $\delta_n^\pm$ of Eq.~\eqref{eq:DLLgap}:
first, as we discussed in the previous section, it must be large enough to accommodate the increasing energy of the sub-levels 
occupied by $N$ particles,
which, in terms of the confinement splitting $\Delta$ and the filling factor $\nu$, means 
$\delta_n^\pm \gg N\Delta /\nu$ (see Eq.~\eqref{eq:eps});
second, the typical interaction strength should be small compared to the gap, 
i.e., $v \ll \delta_n^\pm$.
These two conditions therefore suggest to fix the value of $q$ around the ones
maximising the energy gaps, i.e., $q^2/B = 0.5$ and $q^2/B = 3.8$
in order to have the $n=1$ and $n=2$ DLLs as lowest ones, respectively, 
as found in Sec.~\ref{ssec:DLL} and illustrated in Fig.~\ref{fig:spectrum}.
Under these assumptions the system is mapped into a
gas of single-component atoms subjected to a magnetic field, since
\eqref{chidelta} defines a one-to-one mapping between the canonical
LL $\psi$'s and the eigenstates $\chi$'s. 

Once the above conditions are ensured, we can conveniently treat interactions by
their projections onto the lowest DLL and
the consequent construction of Haldane pseudopotentials (HPs) (see, for example, \cite{jain}).
Let us consider a generic two-body potential
\begin{equation}
V_{ss'} 
  =\frac{1}{2\pi}\int d^{2}k \, V(k) \, 
e^{i\vec{k}\left(\vec{r}_{i}-\vec{r}_{j}\right)}\left|ss'\right\rangle \left\langle ss'\right| \ ,
\end{equation}
rotationally and translationally invariant in the $xy$-plane,
as well as symmetric under spin-exchange $s\leftrightarrow s'$,
as it will indeed be the case for our fermionic mixture in Eq.~\eqref{eq:interaction}.
For the moment, let us anyway keep the discussion independent of the statistical nature of the particles. 
We adopt here the so-called {\it disk} geometry, since it is the one
more immediately linkable to the experimental setup.

The lowest DLL projected eigenstates of $V_{ss'}$ are two-body states with a well-defined
relative (azimuthal) angular momentum $m_{\rm rel}$.
The corresponding matrix elements are then 
\begin{multline} \label{eqwn}
W_{m_{\rm rel}}^{(n)} =
    V_{m_{\rm rel}}^{n-1,n-1} \cos^{4}\frac{\varphi_{n}}{2} 
+ V_{m_{\rm rel}}^{n,n} \sin^{4}\frac{\varphi_{n}}{2} + \\
+ V_{m_{\rm rel}}^{n,n-1} \frac{\sin^{2}\varphi_{n}}{2}
\, ,
\end{multline}
where we made use of the mapping~\eqref{chidelta} between the canonical LL $\psi$'s
and the effective DLL $\chi$'s.
The three terms correspond respectively to the three possible spin states 
$\ket{\uparrow\uparrow},\ket{\downarrow\downarrow},\ket{\uparrow\downarrow}$.
The coefficients $V_{m_{\rm rel}}^{n,n'}$ are given by the projection of the interaction
on the canonical LL $\psi_{n}$'s of Eq.\eqref{eq:LL}: 
\begin{equation} \label{eqvn}
V_{m_{\rm rel}}^{n,n'}=\int k\, dk\, V(k) L_{n}\left(\frac{k^{2}}{2}\right)L_{n'}\left(\frac{k^{2}}{2}\right)L_{m_{\rm rel}}\left(k^{2}\right)e^{-k^{2}},
\end{equation}
where $L_{p}$ is a Laguerre polynomial.
Let us notice here that, since we are dealing with an effective single-species gas on the lowest DLL,
only odd $m_{\rm rel}$ values will be relevant for fermions and only even $m_{\rm rel}$ for bosons.

\begin{figure}
 \includegraphics[width=\columnwidth]{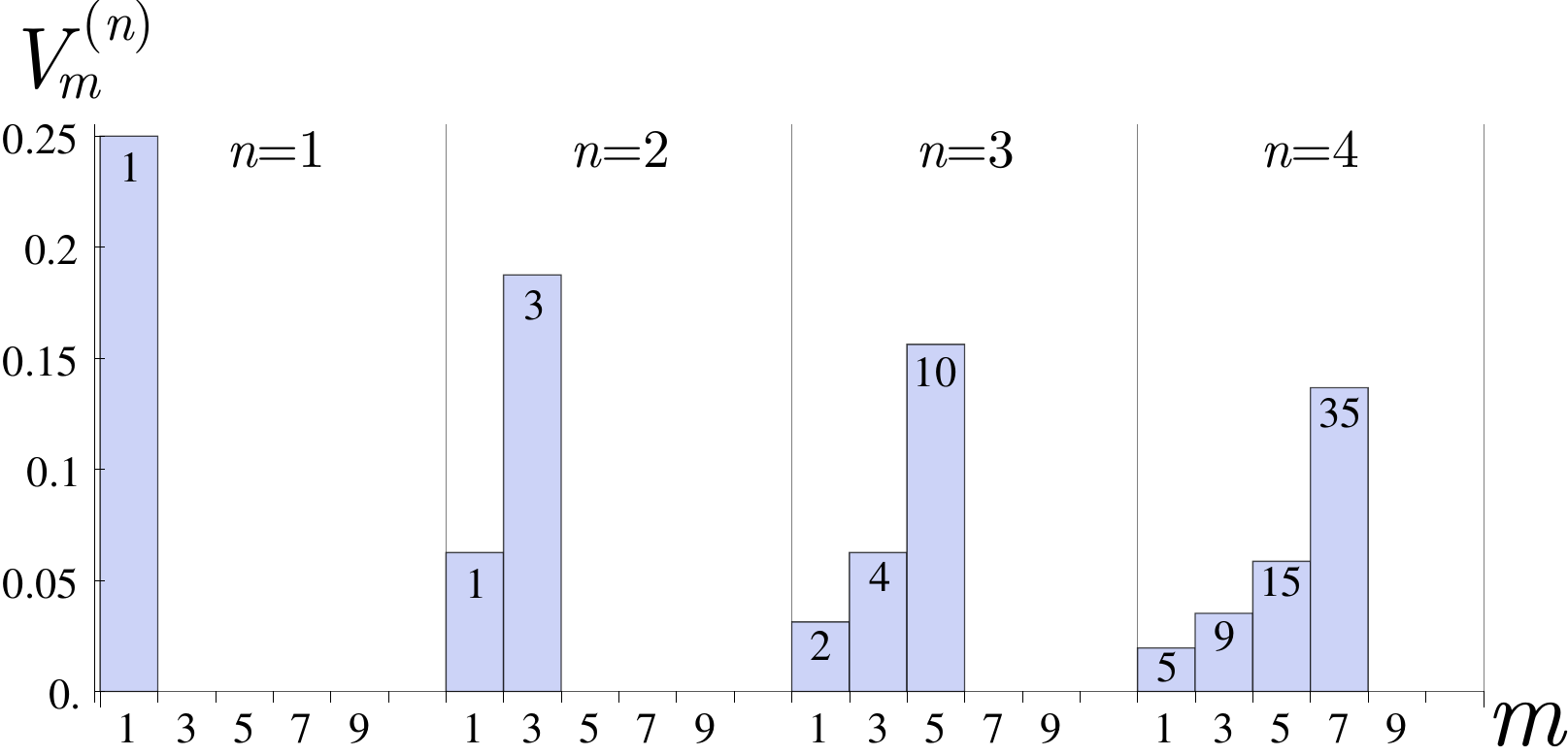}
 \caption{The coefficients $V^{(n)}_m=v\cdot \tilde{v}^{(n)}_{m} / 2^{2n} $, related to the HPs by $W^{(n)}_{m}=V^{(n)}_m \sin^2\left( \varphi_n\right) /2$, are plotted for $v=1$ and $n=1,2,3,4$. The integer numerators $\tilde{v}^{(n)}_m$ are specified in the chart. Only fermionic odd relative angular momenta $m$ are considered. $V^{(n)}_m$ increases with $m$, and then drops to zero for $m\ge 2n$. \label{tab:Wn}}
\end{figure}

For an ultracold fermionic mixture, the effective $s$-wave contact potential 
reads
\begin{equation} \label{eq:interaction}
\hat{V} = v \sum_{i<j}\delta\left(z_{i}-z_{j}\right)
\ket{\uparrow\downarrow}\bra{\uparrow\downarrow}.
\end{equation}
Due to the absence of polarised scattering channels, the first two addends of Eq.~\eqref{eqwn} are null,
and the relative importance of the HPs $W_{m_{\rm rel}}^{(n)}$
is therefore fixed by the canonical cross-coefficients $V_{m}^{n,n-1}$ only. 
These can be then obtained by substituting $V(k)=v/2\pi$ into Eq.~\eqref{eqvn}
and exploiting the properties of the Laguerre polynomials. 
In particular, $V_{m}^{n,n-1} = 0 \ \ \forall m > 2n-1$ and $V_{m}^{n,n-1}~=~V_{2n-1-m}^{n,n-1}$,
and they can be written in the form $V^{(n)}_m = v / 2^{2n} \cdot \tilde{v}^{(n)}_{m}$
with $\tilde{v}^{(n)}_{m}$ an integer number. 
These sequences for the first four regimes of lowest DLL are reported in Fig.~\ref{tab:Wn} 
for the fermionic odd $m_{\rm rel}$'s~\cite{burrellothesis,cooper12}.

It is immediately evident that the HPs behave in a radically different way from the common situation
in solid state devices (like GaAs heterostructures), where they quite rapidly decay with $m_{\rm rel}$.
Instead, they here first grow considerably up to a threshold $m_{\rm rel} \leq 2n-1$,
and then they suddenly drop to exactly zero: henceforth, we dub this non-monotonic behaviour as NMHP.
Let us also notice that a bosonic gas subjected to the same two-body potential 
would experience effective HPs that are monotonically decreasing.
In particular the mapping $m \leftrightarrow 2n -1 -m$ provides the results for interspecies interactions. Intraspecies interactions instead involve all the terms in \eqref{eqwn}, thus introducing further non-zero HPs for $m_{\rm rel}=2n$ \cite{burrellothesis,cooper12}, but without changing the monotonic behavior.
For example in the first DLL, intraspecies interactions are characterized by both $W_2$ and $W_0$ and one can find the constraint $W_2/W_0<2/11$. 

In order to get similar NMHPs in bosonic systems, one 
would need to  consider instead
dipolar/long-range interactions combined with attractive contact potentials, as studied in \cite{grass13},
where the ratios between $W^{(n)}_{m_{\rm rel}}$ can then be tuned.
The enhancement of higher pseudopotentials 
in a non-monotonic way (henceforth NMHP)
out of purely contact interactions (i.e. a single canonical HP) may be then considered
a hallmark of a fermionic mixture in presence of a $U(2)$ gauge potential with strong spin-orbit coupling.%

The above analysis and the values of the HPs in Fig.~\ref{tab:Wn} bring us to the conclusion that, 
for $q<q_{1}$, we have to recover the known results
for the hard-core model with a single non-zero pseudopotential in
fermionic systems (see, for example, \cite{regnault04,cappelli98}). 
In particular this shows that, despite the fermionic nature of the atoms and their behaviour being effectively spin-polarized, a contact repulsion in the presence of the spin-orbit coupling is enough to have a non-zero effective interaction able to reproduce, for example, the Laughlin state at filling $1/3$.
For $q_{1}<q<q_{2}$, instead, due to the presence of NMHPs, 
we may expect the appearance of new strongly correlated states.
Before digging into the investigation of these two regimes 
in Secs.~\ref{below} and \ref{above}, respectively,
we pause here to provide details on the effect of trapping and 
on the second quantized form of the 
Hamiltonian we actually use in the numerical calculations.

\subsection{Effect of the trap on the Haldane pseudopotentials} 
\label{ssec:trapHP}

The presence of a (small) trapping coefficient $\Delta$ is reflected into a correction to the angle $\varphi_n$,
according to 
\begin{equation}\label{overall}
 \sin^2\varphi_n =\frac{8q^2Bn}{8q^2 B n + (B-\frac{\Delta}{2})^2} \ .
\end{equation}
It will therefore affect the precise numerical value of the HPs 
but not the ratio between their angular components,
(see Fig.~\ref{tab:Wn}), which is the distinctive character 
of the physical setup studied here. 
The relative variation of the pseudopotentials $W^{(n)}_{m_{\rm rel}}$ 
will be of order $\Delta/ (2 (B + 8 q^2 n))$,
i.e. a few $10^{-3}$ for the above mentioned regime of 
$\Delta / {\cal B} \simeq 0.005 \div 0.05$. 
As already discussed around Eq.~\eqref{qcross}, 
the confinement also slightly changes the position of DLL crossings, and 
therefore gives minor corrections to the optimal working points
for the lowest DLL approximation.

As a consequence, the main contribution of the 
trapping to the eigenenergies of the interacting system 
is just their shift by $-\Delta L_z$, which will determine the sequence of 
stabilised incompressible 
many-body ground states discussed in the following Sections.
We also stress here that, when we will vary the trap frequencies in our 
numerical calculations,
we actually vary $\Delta$ by keeping fixed the physical synthetic field 
${\cal B}$
(therefore slightly changing the total $B$),
which is indeed the one to be kept  constant in an experimental implementation of Hamiltonian \eqref{ham2}.

\subsection{Exact diagonalization procedure} \label{ssec:exactdiago}

The numerical analysis presented in the next Sections
is based on the exact diagonalization of the interacting Hamiltonian
projected onto the lowest DLL.
It is convenient to treat it in second quantization, 
in terms of the operators $a_{n,m}^{\dag}$ and $a_{n,m}$,
which create or annihilate a fermion in the state $\chi_{n,m}$.
Since we are always working within a single DLL,
we can safely drop the level index $n$ and
treat the index $m$ as labelling positions on a 1D chain.
As usual, we can account for the fermionic anti-commutation relations
$\{ a_{m}, \, a_{m^\prime}^{\dag} \} = \delta_{m,m^\prime}$
via a Jordan-Wigner mapping, and manage the Fock space configurations
as states of a spin $1/2$ chain.

Due to the rotational invariance of the single particle 
Hamiltonian~\eqref{ham2} and of the two-body 
potential~\eqref{eq:interaction}, the total (azimuthal) angular momentum
\begin{equation}
L_{z} = N n - L_{{\rm tot}} \equiv N n -  \sum_{m} m a_{m}^{\dag} a_{m}
\end{equation}
is preserved (as well as the total number of particles $N$).
Therefore the trapping term $\Delta \, L_{\rm tot}$ commutes with the rest of the Hamiltonian, 
which has to be diagonalized in each sector of $L_{\rm tot}$.  
The only non-diagonal part of the Hamiltonian is the interaction $V$ of ~\eqref{eq:interaction}, that involves 
the scattering of a pair of particles from incoming angular labels 
$(j_{1}, \, j_{2})$ into outgoing $(j^\prime_{1}, \, j^\prime_{2})$,
with the selection rule $j_{1}+j_{2} = j^\prime_{1}+j^\prime_{2} \equiv M$.
The initial and final wavefunctions have to be expressed in the 
center-of-mass reference frame,
i.e. in terms of their relative angular momentum $m_{\rm rel}$,
through the coefficients $g\left[m,M,j\right]$ computed in the Appendix 
\ref{app_dec}.
Once this is done, the interaction coefficients are the pseudopotentials 
$W_{m_{\rm rel}}^{(n)}$ 
(calculated in the previous Section), containing all the 
information about the actual wavefunction of the particles
(here in the disk geometry).

In view of these considerations, the Hamiltonian reads 
\begin{widetext} 
\begin{equation} \label{twobody}
H_{int} = \sum_{M} \sum_{m_{\rm rel}} W_{m_{\rm rel}}^{(n)} \sum_{m_{1},m_{2}} 
g\left[m_{\rm rel},M,m_{1}\right] g\left[m_{\rm rel},M,m_{2}\right] 
a_{M-m_{1}}^{\dag} a_{m_{1}}^{\dag} a_{m_{2}} a_{M-m2} \, .
\end{equation}
\end{widetext} 
This will be diagonalised in each separate angular momentum sector $L_{\rm tot}$,
in order to plot the so-called {\it yrast} spectrum.
Through that we can read out 
the appearance of fractional quantum Hall states,
namely by their property of being gapped and incompressible.
First, a state is gapped if it is separated in energy from excitations with the same $L_{\rm tot}$
by an energy difference $\delta E\left(L_{{\rm tot}}\right)$.
Second, a state is incompressible if a certain finite energy cost 
is required to decrease its 
angular momentum, i.e., if the difference in energy to the ground state 
with $L_{\rm tot} - 1$ is non-zero.
Indeed the canonical orbital wavefunctions $\psi_{n,m}$ of Eq.~\eqref{eq:LL} 
(and therefore the deformed $\chi_{n,m}$)
have a mean-squared radius that increases quasi-linearly with the angular 
label $m$: 
decreasing $L_{\rm tot}$ corresponds to increasing the particle density 
in real space.
We can therefore define an incompressibility gap (measured in absence of trapping, $\Delta = 0$)
that summarises these two properties:
\begin{equation} \label{incompr}
\mathcal{D}\left(L_{{\rm tot}}\right)=\min \left( 
\delta E\left(L_{{\rm tot}}\right),E\left(L_{{\rm tot}}\right)-E\left(L_{{\rm tot}}-1\right)
\right)\, .
\end{equation}
We recall that, in the disk geometry adopted here, 
a useful simplification holds~\cite{jain}:
$\mathcal{D}=\delta E$ whenever $E\left(L_{\rm tot} \right) \neq E\left(L_{\rm tot} -1 \right)$. 
Once the trapping $\Delta$ is turned on, the energy penalty $\Delta L_{\rm tot}$
forces the overall ground state of the system to be one of the gapped incompressible ones,
identified by $\mathcal{D} > 0$ and often dubbed as cusp states.

Further details on the numerical algorithm can be found in Appendix 
\ref{app_num}.

\section{Incompressible states in the first deformed Landau level}\label{below}

We present here the results obtained for a spin-orbit coupling $q^2 = 0.5 B$, 
which represents the optimal tuning point for the $\chi_1$ DLL 
to serve as lowest one.
According to Fig.~\ref{tab:Wn}, the only nonvanishing pseudo-potential is 
$W^{(1)}_1 = v \sin^2 \varphi_1 / 8 \simeq 0.1 v \, (1 + 0.1 \Delta)$. 
The physics is therefore expected to reproduce known results of the spinless hardcore model
(see, for example~\cite{regnault04,cappelli98}), 
the only remarkable difference being visible in the spin-resolved density 
profiles (as we illustrate below).
We therefore use this Section as a kind of cross-check for our algorithm,
before sailing off to the unknown in the following Sec.~\ref{above}.

\begin{figure}[tb]
\includegraphics[width=1\columnwidth]{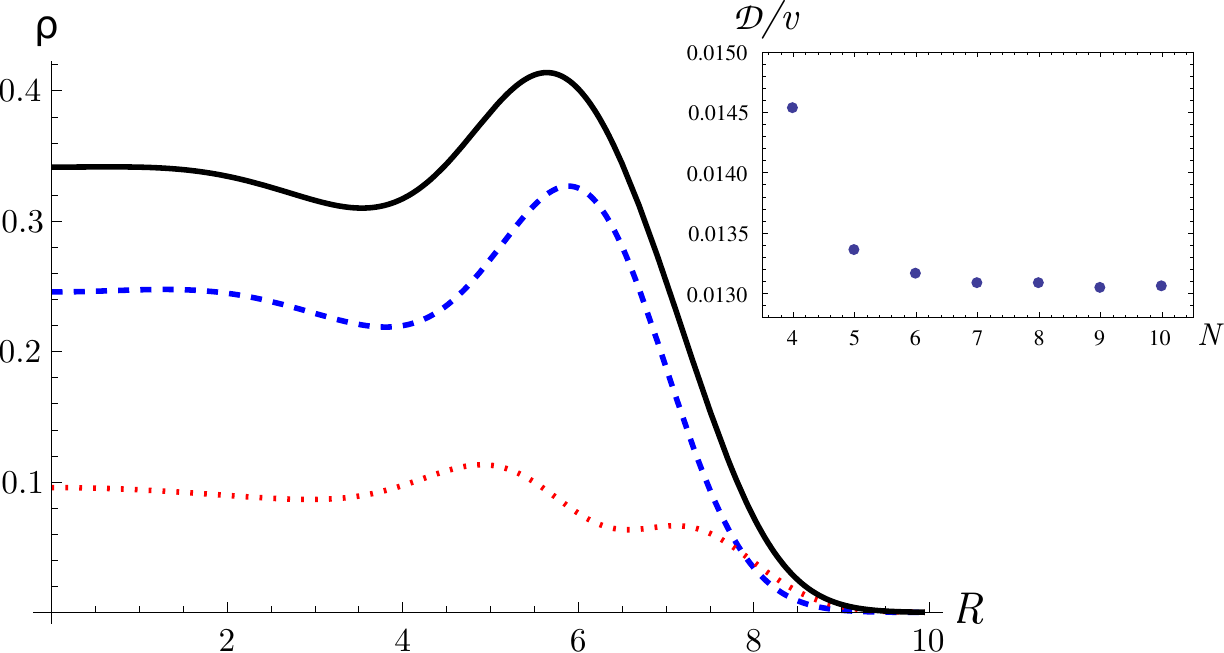} 
\caption{Radial density profile for the up (dashed blue line) and down (dotted red
line) components of the Laughlin state with $10$ 
atoms at $L_{{\rm tot}}=135$ in a system with $q^{2}=B/2$.
The total radial density is plotted as a continuous black line. 
Inset: Gap $\mathcal{D}$, in units of $v$, for the Laughlin state at filling $1/3$ as a function of $N$.}
\label{laughlin} 
\end{figure}

The sequence of cusp states starts with the Laughlin state at $\nu = 1/3$, i.e. $L_{\rm tot} = L_{\rm Lau}= 3 N (N-1) / 2$,
which can be seen as the densest one with zero interaction energy
or, viceversa, as to the incompressible state with largest angular momentum. 
Its total and spin-resolved radial densities
are plotted in Fig.~\ref{laughlin} for the case with $N=10$ particles (the total radial density is normalized such that $\int \rho(r) r dr=N$). 
The overall distribution resembles a flat disk with height $1/3$, as in the standard case,
whereas the $\up, \down$ spin components are not simply rescaled by a factor
$\cos^2 \varphi_1$ and $\sin^2 \varphi_1$, as naively expected from the
superposition coefficients.
The radial distribution of the canonical $\psi_{0,m}$ determining the $\up$ profile
are indeed different from the ones of $\psi_{1,m}$ ruling over the $\down$ density,
in a non-uniform way with $m$.
In particular, the $\up$ component has a more pronounced peak and a sharper edge,
while the $\down$ is more smoothly washed out and slightly prevails in the outmost regions.

\begin{figure}
\includegraphics[width=1\columnwidth]{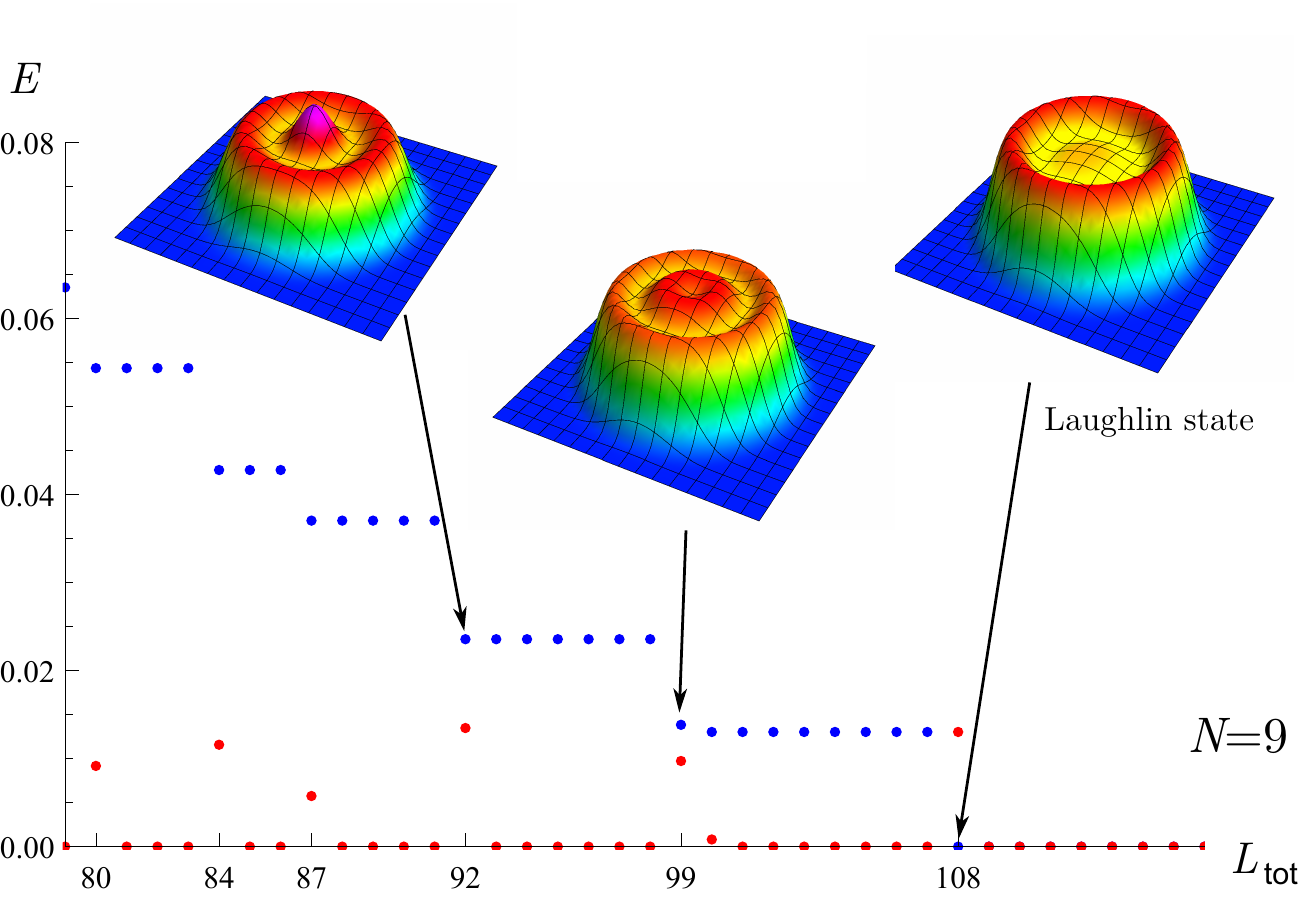} 
\caption{The low-energy part of the yrast spectrum (blue) 
and the related compressibility gap (Red) are plotted 
as a function of the total angular momentum
$L_{{\rm tot}}$ for $N=9$ atoms. The energies are expressed in units
of the interaction constant $v$ ($\Delta=0$).
On the horizontal axis we reported the angular momenta of the main
incompressible states. $L=108$ corresponds to the Laughlin state,
$L_{{\rm tot}}=99,92,87,84$ correspond respectively 
to the presence of $1,2,3$ and $4$ quasiparticles.
In the insets the density profiles of the states at $L_{{\rm tot}}=108,99$ and
$92$ are plotted for $q^{2}=B/2$. The Laughlin state presents an
almost flat plateau, whereas the states at $L_{{\rm tot}}=99$ and 
$L_{{\rm tot}}=92$ show
the typical profile of an odd or even number of quasiparticles in
the center of the system.}
\label{yrast9} 
\end{figure}

\begin{figure}
\includegraphics[width=1\columnwidth]{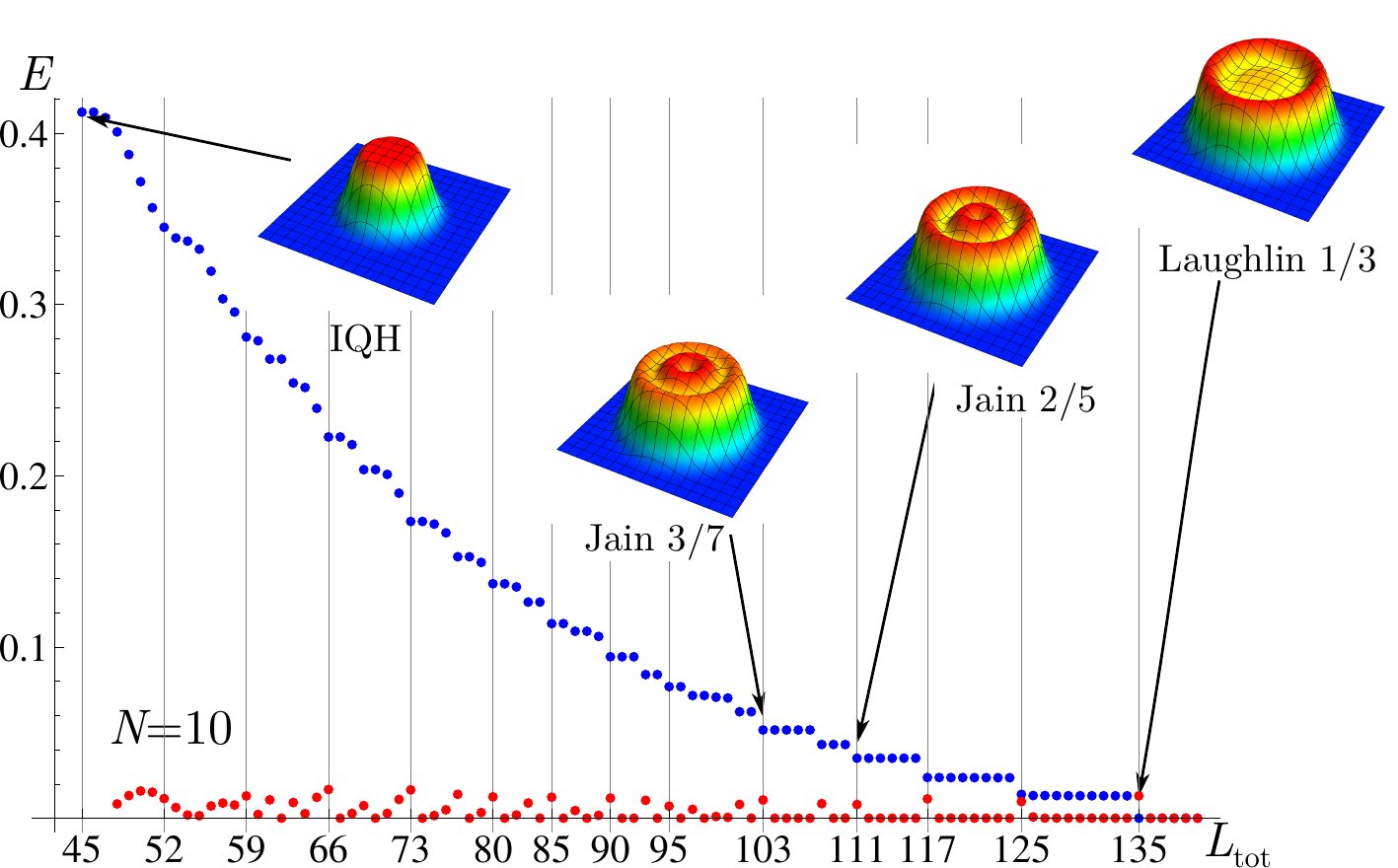} 
\caption{Full yrast spectrum (blue) and related compressibility gap (red) for
$N=10$ fermions. In the insets the total density profile of the IQH 
state, of the candidate Jain states at filling
$3/7$ and $2/5$ and of the Laughlin state are plotted for $q^{2}=B/2$
and $\Delta=0$. On the horizontal axis the momenta of the states
stabilized by the trapping potential are reported.}
\label{yrast10} 
\end{figure}

\begin{figure}[htb]
\includegraphics[width=1\columnwidth]{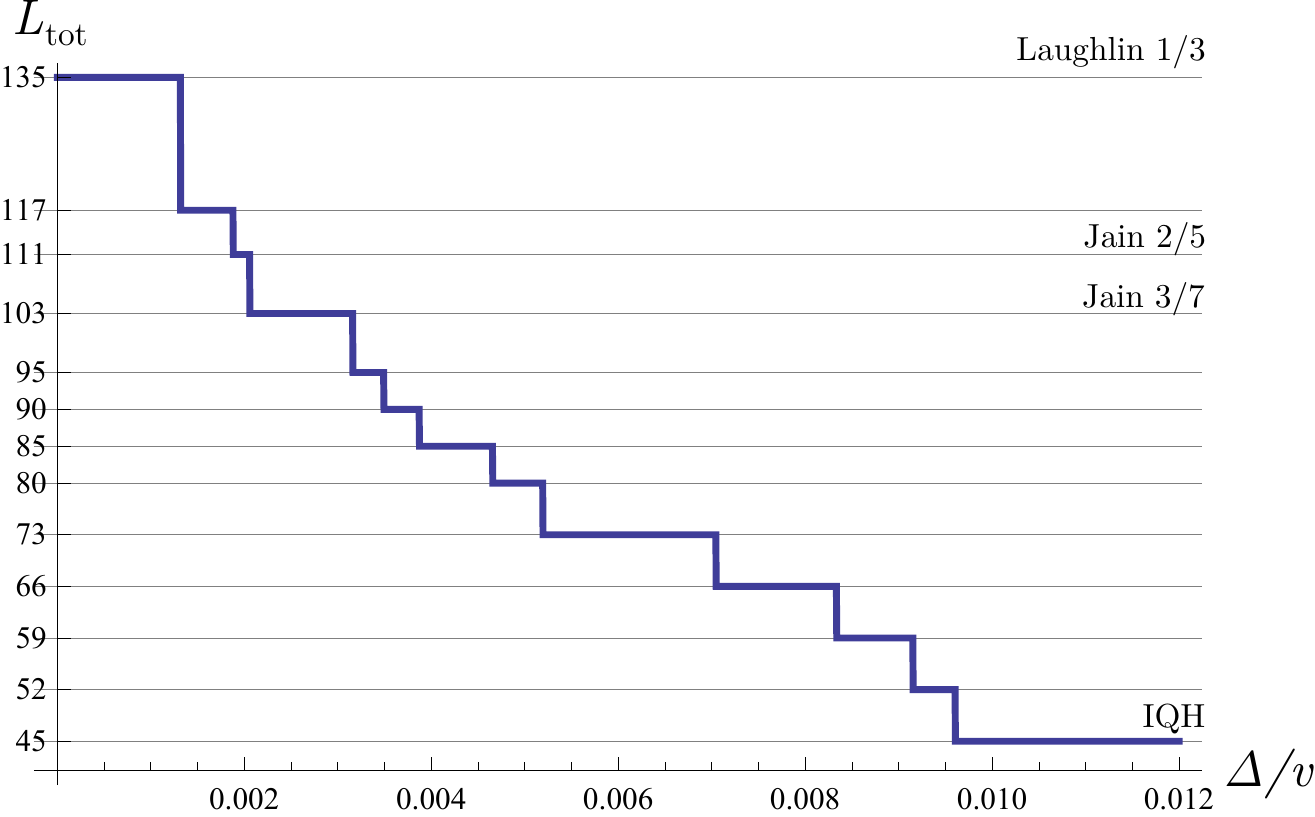} 
\caption{Angular momentum of the ground state for $N=10$ fermions as a function
of the trapping parameter $\Delta/v$,
keeping constant the ratio between $q^2$ 
and the magnetic field ${\cal B}$ at the value $q^2/{\cal B}=0.5$ in the first 
DLL (with ${\cal B}=1$). The horizontal lines define the incompressible
ground states stabilized by the trapping potential. 
These cusp states include the two candidate Jain states identified in \cite{cappelli98}.
For higher densities several ground states alternate, consistent with eightfold rings with two fermions in the center \cite{emperador05}. 
}
\label{trap10} 
\end{figure}

The yrast spectra for $N=9$ and $N=10$ are shown in Figs.~\ref{yrast9}-\ref{yrast10}, respectively,
while in Fig.~\ref{trap10} we show the sequence of ground states that are selected
by the trapping $\Delta L_{\rm tot}$.
By decreasing the angular momentum from the Laughlin state,
several plateaus appear, characterized by decreasing lengths $N,~N~-~2,~N~-~4,~\ldots$
and separated by each other by approximately the same gap
$E_{\rm qp} \simeq \mathcal{D}(L_{\rm Lau}) \approx 0.013 v$. 
This corresponds to the creation of a quasiparticle in the center of the system,
as visible also from the density profiles shown in the insets of Figs.~\ref{yrast9}-\ref{yrast10}:
each plateau represents then a branch of states with the same number of bulk quasiparticles. 
Both the length and the separation of these plateaus can be successfully understood by
the composite fermion model~\cite{jain92,cappelli98,jain07},
which describes them in terms of integer quantum Hall (IQH) states
of composite fermions populating a set of effective LLs, called $\Lambda$-levels.
The $\Lambda$-levels are separated by $E_{\rm qp}$,
which is acquired each time a particle is promoted on the next level to decrease the angular momentum.
If one furthermore constrains the composite fermions to occupy only the first two  $\Lambda$-levels,
the sequence $L_{\rm Lau} - p (N+1-p)$, $p \in \mathbb{N}$, is recovered.

By increasing the energy of the system, however, not all the states predicted
by the composite fermion model appear in the spectrum~\cite{cappelli98}
and the pattern of the quasiparticle states is altered.
For example, for $N=9$ (Fig.~\ref{yrast9}, $L_{{\rm Lau}} = 108$)
we are able to distinguish four such plateaus  at $L_{{\rm tot}}=99, 92, 87, 84$,
that exactly match the model predictions,
while for $N=10$ (Fig.~\ref{yrast10}, $L_{{\rm Lau}} = 135$) only three such structures
are visible at $L_{{\rm tot}}=125, 117, 111$.
Moreover, a residual interaction among the composite fermions
gives rise to some sort of pairing effects, 
that make the creation of pairs of quasiparticles more convenient,
as visible also in our yrast spectra
and in the absence of $L_{{\rm tot}}= 125$ in Fig.~\ref{trap10}. We verified that a similar behavior appears also for $N=9$.
The presence of a characteristic density peak in the center of the system
also constitutes a detectable signature of such pairing.

Due to the finite number of particles, it may become quite rapidly difficult to univocally identify 
the nature of the incompressible gapped states with decreasing $L$.
An example is provided by the states with $N=10$ particles at $L_{{\rm tot}}=111$,
whose momentum corresponds to three composite fermions in the second $\Lambda$-level,
therefore to  three bulk excitations over the Laughlin.
However, it was also considered as a candidate for the Jain state at filling $2/5$~\cite{cappelli98,macdonald03},
i.e., roughly speaking the full occupation of the first two $\Lambda$-levels, aside from boundary corrections.
The analysis of the edge excitations seems to privilege such a Jain picture over the Laughlin one~\cite{macdonald03}.
By further scanning the yrast spectrum of Fig.~\ref{yrast10}
(or equivalently, by increasing the trapping in Fig.~\ref{trap10}),
the bulk excitation picture definitely breaks down.
An interesting stable state is then found at $L_{{\rm tot}}=103$,
which seems consistent with a Jain-state at filling $3/7$
with composite fermions fully occupying also the third $\Lambda$-level \cite{cappelli98}.

The energy gap related to such candidate Jain states has been shown to be non-vanishing
in the thermodynamic limit by previous studies in other geometries (both the torus and the sphere)%
~\cite{regnault04}.
Therefore, it seems plausible that the trapping allows
a transition from the Laughlin quasiparticle regime to the Jain sequence,
once the parameter $\Delta \sim \omega^2 / 2\mathcal{B}$ 
exceeds a threshold of order $E_{\rm qp}/N \approx 0.013v / N$,
as already suggested in \cite{regnault04}.
Conversely, a similar estimate holds for the whole fractional Hall regime,
and explains one of the main difficulties that have so far made difficult 
its realization
in cold atomic samples with strategies based on the sole rotation of the sample~\cite{cooper08,fetter}.

Instead of discussing the yrast at even lower $L$'s,
we rather prefer to examine only the states that are stabilised by the presence of
the harmonic trapping $\omega$ (see Fig. \ref{trap10}). 
This leads to a quite regular pattern of ground states, equally spaced in angular component $L$
($95$ to $80$ by $5$, and $80$ to $45$ by $7$),
down to the maximally dense integer quantum Hall state, $\nu=1$, at $L_{\rm tot} = N (N-1)/2 = 45$.
Rather than bulk excitations of the previously built Jain states, 
they can most probably be interpreted as less correlated states characterized
by peculiar geometrical distributions of the atoms.
A striking example is the $L_{\rm tot} = 85$ state,
which at first sight might be tempting to relate to the Moore-Read Pfaffian state at $\nu=1/2$:
its density profile presents instead a pronounced peak of two particles in the centre surrounded
by an eightfold ring.
This configuration is typical of similar models of quantum dots
in the fractional quantum Hall regime~\cite{emperador05}:
therefore, we confirm the well-known result that, differently from the bosonic case \cite{cooper01},
the Pfaffian state does not appear among the ground states of the fermionic hard-core model.

In Appendix \ref{app_est} we find the approximated critical value $\Delta_c^{(1)}=v/(5\pi N)$ to provide a rough estimate of the trapping threshold between the intermediate and the regular densest states culminating in the integer quantum Hall droplet. This results in $\Delta_c^{(1)} \approx 0.0064v$ for $N=10$ (to be compared with the data in Fig. \ref{trap10}) and we expect the estimate to improve with increasing $N$.

Before concluding this Section, let us observe that 
bosonic atoms would behave in a very similar way in the first DLL. In 
bosonic systems, indeed, we would have two different behaviors for interspecies and intraspecies interactions \cite{grass12b}. In the case of interspecies interactions without intraspecies ones, only the pseudopotential $W_0$ is different from zero. Therefore we recover the known results with the Laughlin state at filling $1/2$ having zero energy and, by increasing the density, quasiparticles and Jain states appear \cite{cazalilla05}. Intraspecies interactions imply that also $W_2$ is different from zero, although $W_2 \ll W_0$. Therefore the interaction energy is zero for the Laughlin state at filling $1/4$ and different from zero for densest states \cite{grass12b}. The study of the incompressible states appearing in this case requires a deeper investigation.

The statistical nature of the underlying system components
will play a radically different role in the second DLL,
due to the fermionic non-monotonic Haldane potentials,
as we are going to discuss next.

\section{Incompressible states in the second deformed Landau level}\label{above}

Here we focus on the regime $q_{1} < q < q_{2}$,
where the lowest DLL is the second one, $\chi_2$:
more precisely, we tune $q^2=3.8 B$, which maximizes the energetic distance from
higher DLLs, as discussed before.
The Haldane pseudopotentials display two distinctive features, as computed in Sec.~\ref{HPs}:
first, they are growing with the relative angular momentum and have a sensibly large
$W^{(2)}_{3} / W^{(2)}_{1} = 3$;
second, they suddenly drop to zero at $m_{\rm rel} > 3$.
This might at first sight remind of a rather unusual situation in more traditional quantum Hall setups,
where non-monotonic residual interactions might arise 
between composite fermions in higher $\Lambda$-levels~\cite{jain}.
There, however, not only does this happen strictly in the dressed picture and not
for the elementary constituents of the semiconducting systems,
but also the HPs do not vanish for large $m_{\rm rel}$'s 
and therefore give rise to an attractive interaction~\cite{jain}.

We present the full yrast spectrum of $N=8$ particles,
up to the most dilute incompressible state, i.e. the Laughlin $\nu=1/5$
at $L_{\rm tot} = 5 N (N-1) /2 =140$ (see Fig.~\ref{yrast8q38}).
We then analyse the incompressible states that are stabilized by the trapping term,
and clearly identify three classes of them
by the different incremental ratio between $L_{\rm tot}$ and $\Delta$ (see Fig.~\ref{trap8}):
a) quasiparticle excitations of the the Laughlin $\nu=1/5$;
b) a series of intermediate strongly correlated states around $\nu =1/3$,
where the Laughlin is ruled out, and a candidate Haffnian state emerges;
c) finally, vortices over the maximally dense integer quantum Hall state $\nu=1$,
i.e.  $L_{\rm tot} = N (N-1) /2 = 28$.
The description of such regimes is presented in the following subsections.
For some specific cases, we present also numerical data for density profiles and correlation functions
with an higher particle number, up to $N=14$.

\subsection{Laughlin $\nu=1/5$ and its quasiparticles}

\begin{figure}
\includegraphics[width=1\columnwidth]{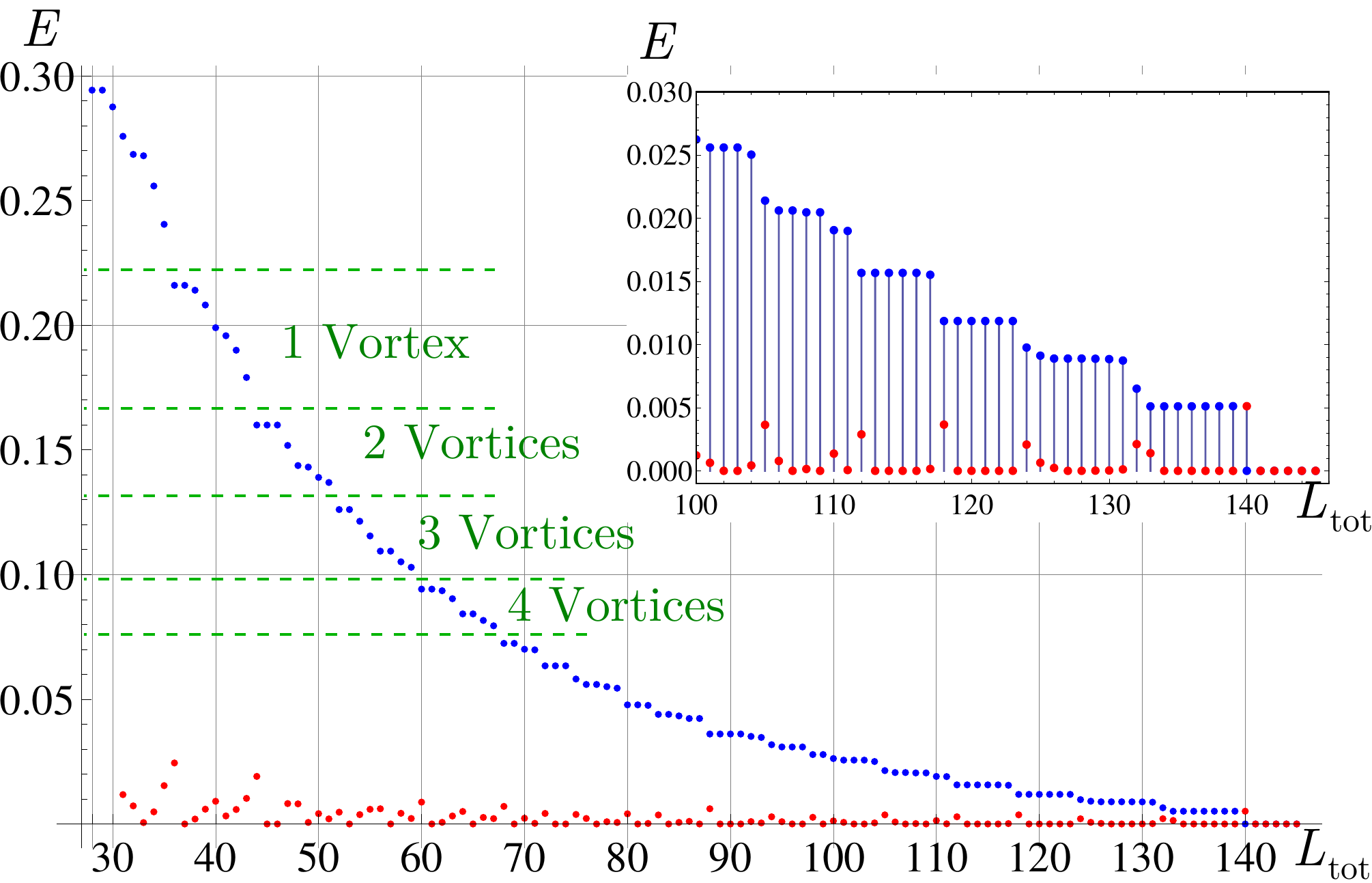} 
\caption{The yrast spectrum (blue) and the incompressibility gap (red) of the
system with $N=8$ atoms is plotted for $q_{1}^{2}<q^{2}<q_{2}^{2}$
as a function of the total angular momentum. The energies are expressed
in terms of the interaction constant $v$. Inset: detail of the lowest
part of the spectrum presenting quasiparticle plateaus.}
\label{yrast8q38} 
\end{figure}

\begin{figure}
\includegraphics[width=1\columnwidth]{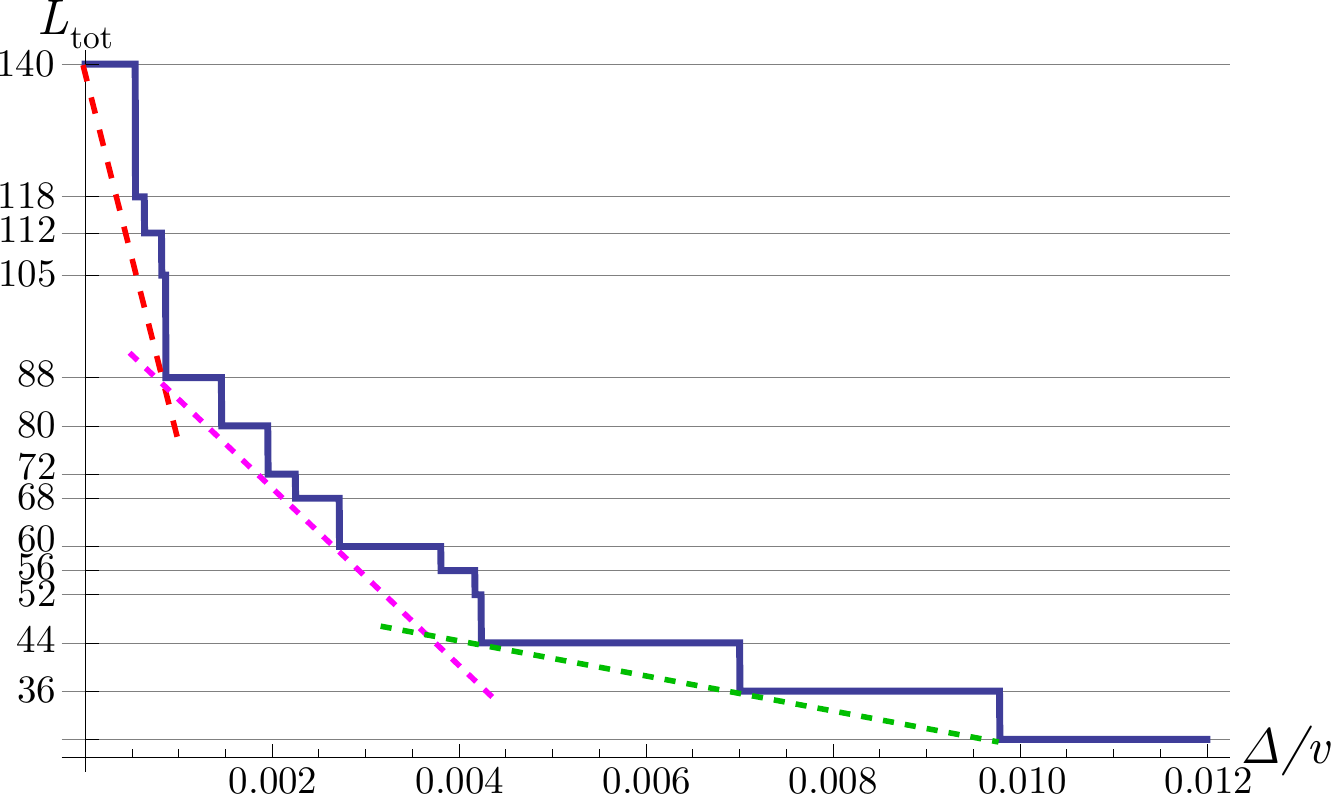} 
\caption{Angular momentum of the ground state for $8$ fermions at
$q^{2}=3.8 {\cal B}$ as a function of the trapping parameter $\Delta/v$
for ${\cal B}=1$. The dotted lines are a guide for the eye
to distinguish the three regimes discussed in the main text.}
\label{trap8} 
\end{figure}

Similarly to what was found in the first DLL, incompressible quasiparticle states
emerge in the lowest part of the yrast spectrum of Fig.~\ref{yrast8q38}
as energy plateaus  at momenta smaller than $L_{\rm Lau} = 5 N (N-1) /2 =140$.
Such plateaus are separated by energy gaps corresponding to the formation of bulk excitations,
considerably smaller (by almost a factor $3$) than the corresponding gap
for the Laughlin $\nu = 1/3$ in the first DLL (see Sec.~\ref{below}).
The transitions among them, however, are smoother than the previous case due to
the effect of the pseudopotential $W^{(2)}_{3}$.
Moreover, the different effective interaction in this regime changes
the angular momenta of the incompressible quasiparticle states,
which do not appear anymore at the values predicted
by the composite fermion construction.
Finally we notice that the overall stability window in $\Delta$
of such Laughlin $\nu = 1/5$ regime is small
when we compare Fig.~\ref{trap8} with Fig.~\ref{trap10}, i.e. when we compare it with the stability of the $\nu=1/3$ Laughlin in the first DLL. However also in semiconductor systems the Laughlin states at low densities are harder to achieve \cite{jiang90}.

\subsection{Intermediate regime $\nu \simeq 1/3$: paired states}

If the trapping is increased beyond the relatively short stability window of the Laughlin $\nu=1/5$ regime,
we can observe a quite sharp change in the incompressibility gaps
and the system enters in a qualitatively different series of strongly correlated states.
Such a series is located in a region roughly compatible with a filling $\nu =1/3$,
i.e. $L_{\rm tot} = 84$ in this case (see Fig.~\ref{trap8}).

\begin{figure}
\includegraphics[width=\columnwidth]{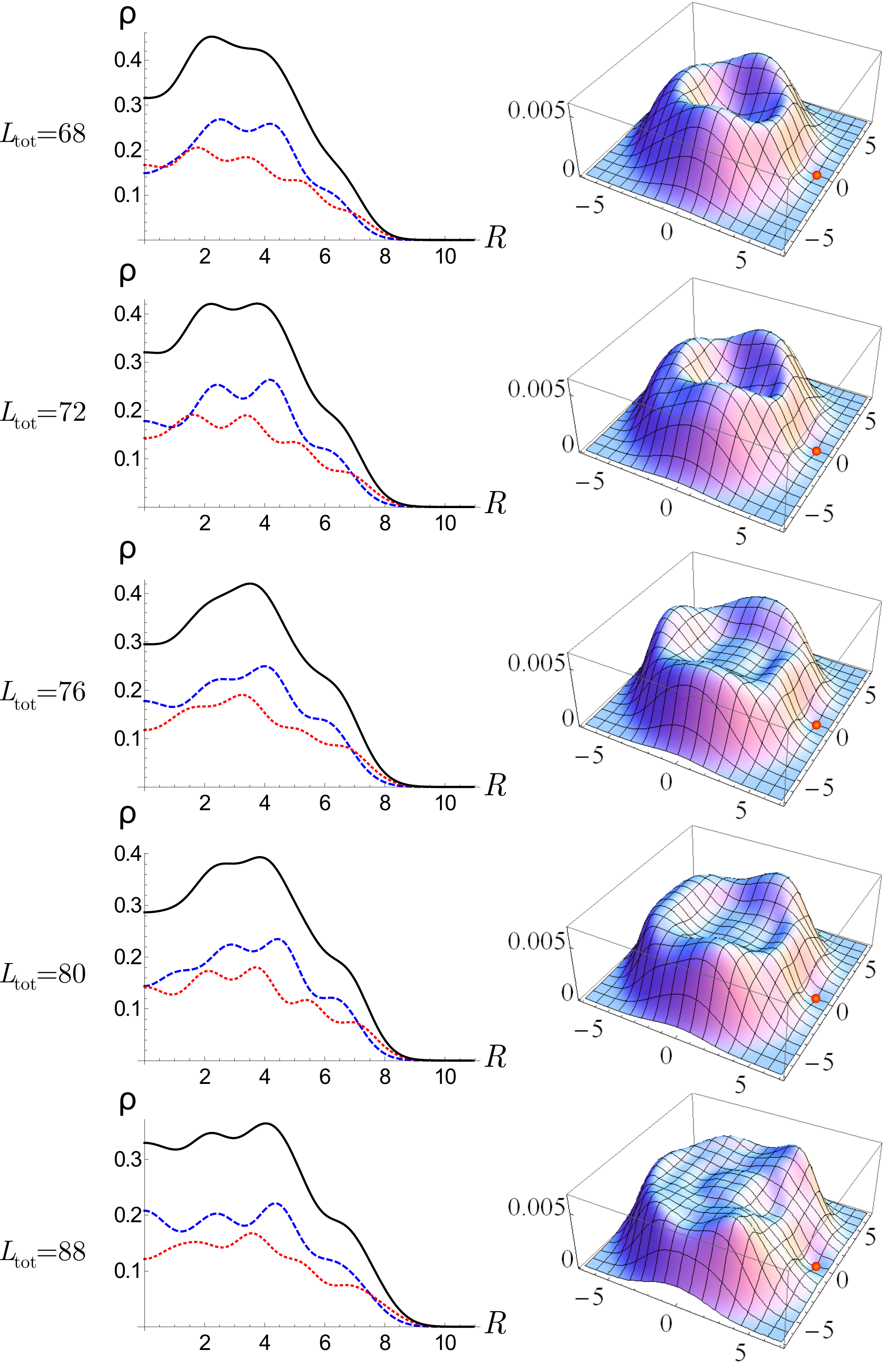}
\caption{Spin-resolved profile densities and effective two-body correlation functions for the 
most relevant gapped states in the intermediate regime for $N=8$. Left: spin up (dashed blue), spin down (dotted red) and total (black) density profiles. Right: two-body effective correlation function with the reference atom in the position $\vec{r}_0=(6,0)$ (red dots).
The states at 
$L_{\rm tot}=68,72,80,88$ are cusp states stabilized by the trap, 
whereas the incompressible state at $L_{\rm tot}=76$ is compatible 
with the Haffnian wavefunction \eqref{haff} and it is showed for comparison. 
The states at $L_{\rm tot}=72,76,80$ show three distinct peaks 
corresponding to pairs of atoms whose presence is consistent 
with the pairing in the Haffnian wavefunction. 
The state at $L_{\rm tot}=68$ may be related to the WYQ state.
The state at $L_{\rm tot}=88$ presents instead an almost flat correlation distribution. 
All the correlation functions are calculated for the single-particle 
wavefunctions $\psi_{0,m}$ instead of $\chi_{2,m}$ 
to make their structure more evident.}
\label{fig:haff}
\end{figure}

The first of the stable cusp states found in Fig.~\ref{trap8},
namely $L_{\rm tot} = 88$, exhibits a significative incompressibility gap in Fig.~\ref{yrast8q38}.
All of the states in its yrast plateau (i.e., up to $L_{\rm tot} = 91$)
present a qualitatively flat density profile and correlation functions,
even when spin-resolved (see Fig.~\ref{fig:haff}), which is clearly distinct from the situation for 
Laughlin $\nu = 1/5$ quasiparticle excitations.
Therefore we think that this cusp state might be a good candidate 
for a {\it primitive} FQH state at filling close to $1/3$.

The following cusp states at $L_{\rm tot}=80$ and $L_{\rm tot}=72$
seem to be of a different nature since they present a density minimum in the center,
as visible in Fig.~\ref{fig:haff}.
Their correlation functions suggest that some kind of pairing between the atoms arise, as showed in Fig. \ref{fig:haff}. In particular we plot effective correlation functions obtained by representing them in terms of the canonical single-particle wavefunctions in the lowest LL, $\psi_{0,m}$, instead of the physical wavefunctions $\chi_{1,m}$. The plotted correlations are thus obtained by evaluating:
\begin{multline}
\varrho\left(\vec{r}\right)= \sum_{jklm} \left[\psi_{0,j}^*\left(\vec{r}\right) \psi_{0,k}^*\left(\vec{r}_0\right) \psi_{0,l}\left(\vec{r}_0\right) \psi_{0,m}\left(\vec{r}\right) \phantom{a^\dag_j}\right. \\ \left. \bra{\Psi}a^\dag_ja^\dag_k a_l a_m\ket{\Psi}\right]
\end{multline}
where the operators $a^\dag_i$ and $a_i$ respectively create and annihilate a fermion in the single-particle state $\chi_{1,i}$ and act on the multi-particle ground state $\Psi$. The choice of adopting these effective correlation functions is done to compare our results with the literature about quantum Hall states and to make the appearance of pairing more evident. In particular we fixed the position of the reference fermion in $\vec{r}_0=(6,0)$ in order to break the rotational symmetry of the represented correlations in such a way that the pairing mechanism becomes more evident and implies the appearance of localized peaks in $\varrho\left(\vec{r}\right)$. We considered $\vec{r}_0$ close to the edge of the sample to avoid strong perturbations of the plotted correlations in the bulk due to the Pauli principle.

In the analysis of the yrast spectrum, we notice the wash-out of the Laughlin state at $\nu=1/3$,
as testified by the absence of relevant incompressibility gap at $L_{\rm tot} = 84$ in Fig.~\ref{yrast8q38}
and by the consistent difference in the orbital occupation numbers (not shown here).
The NMHPs, indeed, penalize the pairs of atoms with $m_{\rm rel} =3$
in favour of the $m_{\rm rel} =1$ pairs, which are normally absent in the Laughlin wavefunction.

The appearance of a pairing mechanism through the enhancement of the pseudopotential $W_3$ is not surprising as it has been verified in analogous bosonic systems \cite{cooper07,grass13} with a strong HP for $m_{\rm rel}=2$. In particular, the effective pairing due to the reduction of the constribution of the two-body wavefunctions at $m_{\rm rel} =3$ might be described by the introduction of a so-called Haffnian prefactor in front of the Laughlin wavefunction, yielding to an effective d-wave pairing of the fermions \cite{wen94}. In particular this generates the so-called Haffnian trial wavefunction~\cite{wen94} which reads (see \cite{green} for the bosonic case):
\begin{equation} \label{haff}
 \Psi_{\rm Hf} \sim 
 \mathcal{S}\left(\frac{1}{\left(z_1-z_2 \right)^2 \ldots\left(z_{N-1}-z_N \right)^2 } \right) 
 \prod_{i,j}\left(z_i-z_j \right)^3
\, ,
\end{equation}
where the operator $\mathcal{S}$ symmetrizes the prefactor over all the permutation of the atoms,
and the wavefunction has to be mapped onto the DLL $\chi_2$ through suitable operators. 
The total angular momentum of the Haffnian for $N=8$ particles is $L_{\rm Hf}=76$,
which unfortunately does not correspond directly to a cusp state.
However, the above mentioned cusp states at $L_{\rm tot}=L_{\rm Hf} \pm N/2 = 80$ and $72$
display quite a lot of qualitative features of the Haffnian ansatz.
In particular, as illustrated in Fig.~\ref{fig:haff}, 
the two atoms in each of the $N/2=4$ pairs are pushed close to each other:
this gives rise to three peaks in the density correlation functions,
while the fourth pair is the one of the reference atom and
the two-body correlation vanishes at its position, due to the Pauli principle.

Such an effect is analogous to the bosonic case, where,
in the presence of a large second HP driven by a long-range dipolar potential,
the Laughlin state at $\nu = 1/2$ disappears in favour of  paired strongly correlated states~\cite{grass13}.
In the fermionic case, though, the wavefunction \eqref{haff} is believed to describe an unstable state. It may be obtained as an exact ground state for specific three body interactions \cite{wen94} but it is known that such interaction brings to a ground state degeneracy on a torus geometry which depends on the number of particle, and thus cannot be taken as a trial wavefunction for a system with topological order \cite{hermanns11}. Nevertheless we believe that the d-wave pairing introduced by the Haffnian prefactor may constitute a good physical mechanism to model the physical states around filling $1/3$ that we obtain on the disk geometry. 

\begin{figure}
\includegraphics[width=\columnwidth]{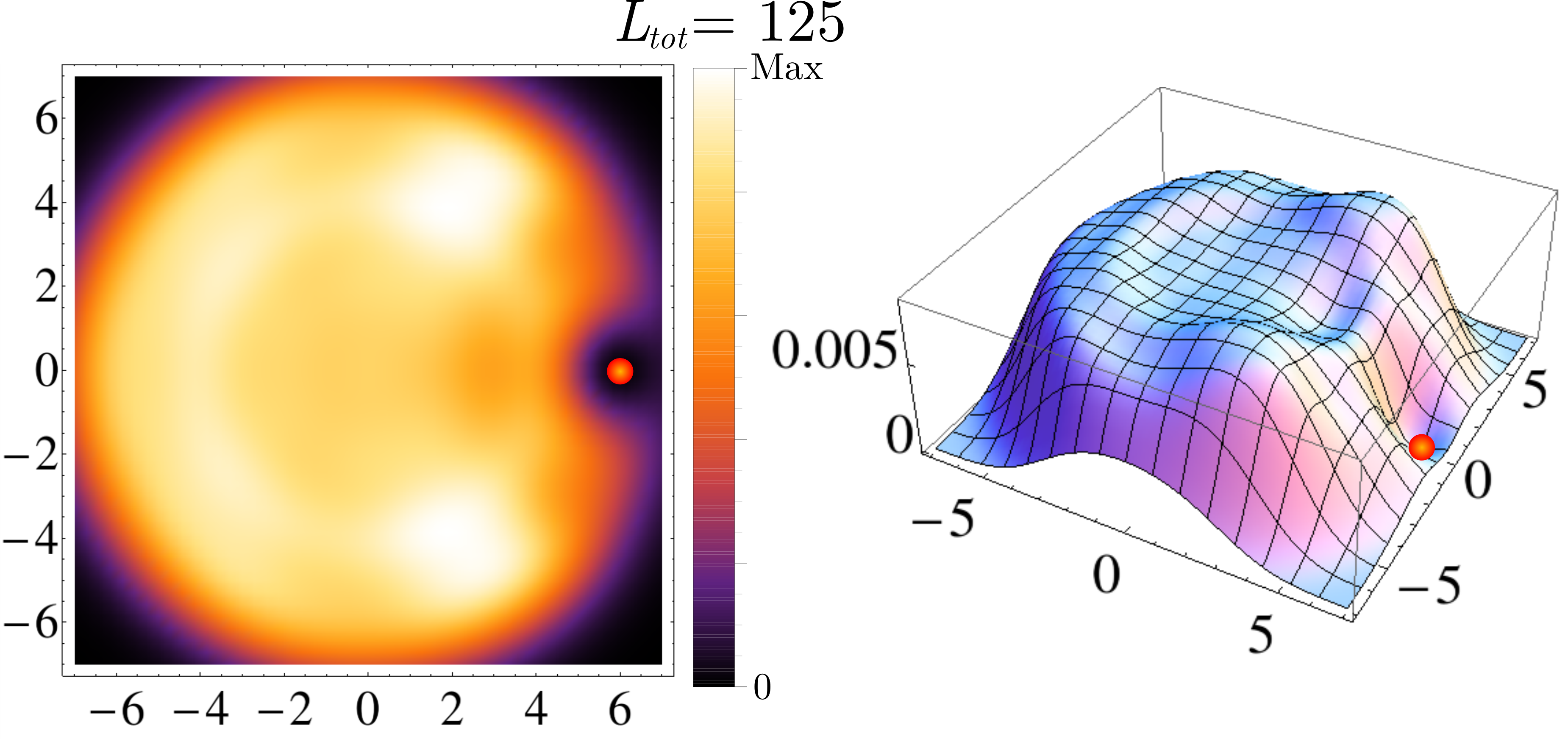}
 \caption{\label{fig:haff_10}
 Effective two-body correlation function of the ground state 
for $N=10$  and $L_{\rm tot}=125$. The red dots represent the position of the reference fermion.}
\end{figure}

The pattern of correlations of the cusp state at $L_{\rm tot}=88$, instead, does not present any clear signature of this pairing mechanism, thus our data suggest a deep  distinction between the {\it paired} states
at $L_{\rm tot}=72, 80$ and this cusp state at $L_{\rm tot}=88$,
which represents then a different sort of fractional quantum Hall state.
The precise characterization of these states however eludes the numerical technique employed here,
as (unfortunately) exemplified by the case of $N=10$ in Fig.~\ref{fig:haff_10}, corresponding to the angular momentum of the Haffnian wavefunction \eqref{haff}. With larger numbers of atoms, the interpretation of the two-body correlation functions becomes progressively more complicated.
In order to better understand the structure of such $\nu \simeq 1/3$ states (and other ones as well),
we proceeded to the study of the entanglement spectrum, presented in Sec.~\ref{ent_sp}.

Finally, by increasing further the density for $N=8$, we find a cusp state at $L_{\rm tot}=68$. This state may be related to the fractional quantum Hall state identified by Wojs, Yi and Quinn (WYQ) in \cite{wojs04} and discussed in relation with the composite fermion state at filling $4/11$ in \cite{wu14}. The WYQ state appears at $L_{\rm tot}=(3N^2-7N)/2$ and it is the ground state for the hollow-core model with $W_1=0$, thus it presents, as expected, a reduced contributions of two-particle wavefunctions at $m_{\rm rel}=3$. However, despite the small $W_1$ HP in our system, it is possible that the WYQ state does not provide a qualitative description of the state we obtain at $L_{\rm tot}=68$. For $N=10$, indeed, the corresponding state at $L_{\rm tot}=115$ results in being compressible, therefore it is likely that the presence of the minor HP at $m_{\rm rel}=1$ is sufficient to drive the system away from the WYQ state. Furthermore, for $N=8$, the incompressible state at $L_{\rm tot}=68$ can also be interpreted as a state describing five vortices in its center, as we discuss in the next subsection.

\subsection{Vortices over the integer quantum Hall state}

\begin{figure*}[ht!]
\includegraphics[width=0.9\textwidth]{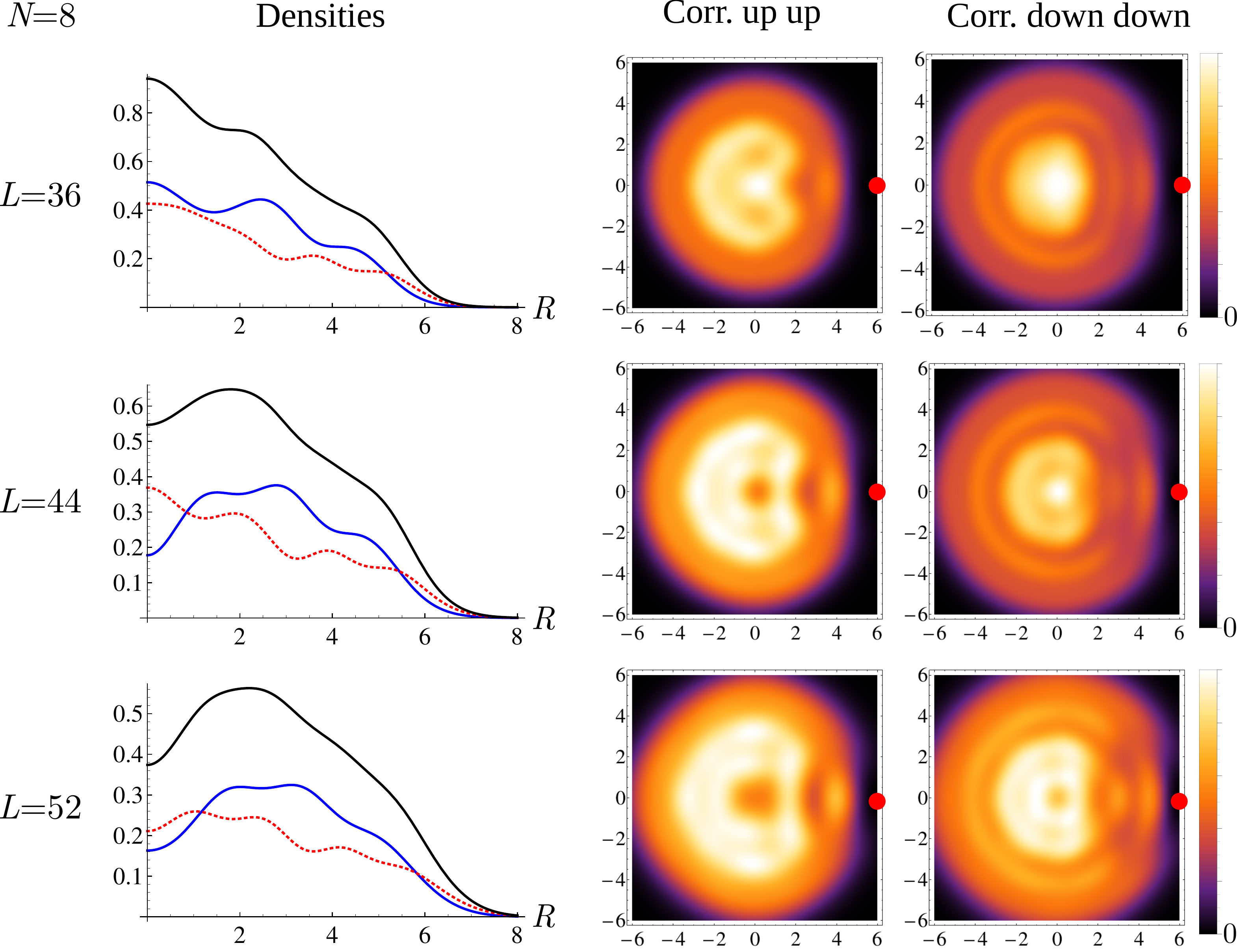}
\caption{\label{fig:vortex}
The density profiles (first column) and the spin 
resolved $2$-body density correlations for the spin up (second column) 
and spin down (third column) component of vortex states 
are showed for a system of $8$ fermions. The considered states 
correspond to one, two and three central vortices, in turn 
corresponding respectively to the cusp states with $L_{\rm tot}=36,44,52$. 
In the first column the blue lines represent the spin up radial density, 
the red dashed lines are the spin down densities and the black ones correspond to the 
total densities. The two-body correlations are plotted in arbitrary 
units fixing the position of one atom in $(6,0)$ (red dots). The insertion 
of the first vortex does not cause a minimum in the 
densities and correlations in the center of the system. 
The second vortex has instead a skyrmionic behavior, 
with minima in the spin up component only, whereas 
the third vortex present a minima in both the up and down components.}
\end{figure*}

By further increasing the trapping, qualitatively different cusp states alternate
as global ground states and the system undergoes a further transition towards a third regime.
This is better understood starting from the maximally dense fermionic droplet,
i.e. the integer quantum Hall state $\nu=1$ at $L_{\rm IQH} = N (N-1) / 2 = 28$,
where all and only the first $N$ orbitals are fully occupied.
It is indeed evident in Fig.~\ref{trap8} that cusp states with quite large energy gaps appear 
every time the angular momentum is changed by $N=8$.
Each jump can be interpreted as the insertion of a vortex in the center of the system,
followed by perturbations on the boundaries, which finally lead to the introduction of a new vortex,
and therefore to the next jump.
Looking at the orbital occupation, indeed, the occupation numbers
$C_{m}=\left\langle a_{m}^{\dag}a_{m}\right\rangle$ 
of the lowest orbitals drop one after the other at each vortex insertion.

The depletion of orbitals with low momentum $m$, however, 
does not immediately translate into a central density dip in real space
(as in the common Abelian case); this is illustrated in Fig \ref{fig:vortex}.
Indeed, the canonical LL wavefunctions $\psi_{1,0}$ and $\psi_{2,0}$, 
involved in the definition of the DLL element $\chi_{2,0}$,
are already vanishing at the center of the trap.
It is therefore the second vortex that qualitatively alters 
the density profile and the spin texture of the IQH state.
The depletion of $\psi_{1,1}$ induces a dip in the $\up$ component, 
leading to a skyrmionic spin texture with the prevalence 
of the $\down$ component at the center of the system
(see also~\cite{estienne11}).
Once the third vortex is introduced,
also the spin down density acquires a minimum in the center (due to $\psi_{2,2}$),
and a pronounced minimum in the total density appears.

The critical value of the trapping parameter $\Delta$ separating the intermediate fractional quantum Hall regime with this vortex phase is roughly predicted by $\Delta^{(2)}_c = v/(8\pi N)$ (see App. \ref{app_est}) which provides a good estimation for the available numerical data $(6 \le N \le 10)$.

By increasing the number of particles,
a task which is numerically not particularly demanding at low momenta $L_{\rm tot}$ (see App.~\ref{app_num}),
the cusp states do not match anymore the description in terms of vortices in the center.
The latter become less and less energetically favourable and they progressively 
disappear from the low energy part of the yrast spectrum,
with the first vortex state at $L=N(N+1)/2$ being no longer a cusp state once $N \ge 13$.
For example, for $N=16$, the integer quantum Hall droplet has $L_{\rm tot}=120$ and we find cusp states at $L_{\rm tot}=134,148$ with jumps different from $N$.
This is consistent with the results of Refs~\cite{saarikoski10,macdonald02} for the Coulomb interaction,
that predict a qualitative change in the vortex behaviour 
for small quantum Hall liquid droplets above $N=12$.
However, due to the axial symmetry of our DLL basis, 
it is difficult to evaluate the rise of more complex geometries of the vortex lattice,
as the ones investigated for fermions in~\cite{saarikoski10}
and for bosonic systems with non-Abelian potentials
in~\cite{cooper12,galitski11,xu10,zhou11,aftalion13}.

\section{Entanglement spectrum}\label{ent_sp}

In order to characterize the properties of quantum Hall states, 
a growing importance has been recently acquired by 
the study of their entanglement features.
Among them, the entanglement spectrum (ES)~\cite{haldane08}
is defined as the set of levels 
\begin{equation} \label{eq:ES}
	{\rm ES}_l \equiv - \ln \lambda_l
\end{equation}
where $\lambda_l$ are the eigenvalues of the reduced density matrix $\rho_A = {\rm Tr}_{\bar{A}} \ket{\psi} \bra{\psi}$ 
of a given bipartition, $S = A \cup \bar{A}$ of the system $S$, and $l=1,2,\dots,\dim{\rho_A}$.

The ES provides rich information on the many-body structure of the state $\ket{\psi}$ under investigation as it allows, for example, to examine its boundary effective theories \cite{ludwig12} or the physics of its hole excitations \cite{bernevig11}.
Several kinds of bipartition have been considered,
the main being the orbital~\cite{haldane08} and the particle ones~\cite{bernevig11},
which offer two complementary perspectives on the quantum Hall states.

The orbital partition consists in selecting a subset $A$ of orbitals within the LL~\cite{haldane08}:
\begin{equation} \label{eq:OES}
	\rho_{A}^{(O)} = \sum_{\{\vec{n}\}_A} \sum_{\{\vec{n}^\prime\}_A} 
		\left( \sum_{\{\vec{n}^{\prime\prime}\}_{\bar{A}}}  \Psi^*_{\vec{n} \otimes \vec{n}^{\prime\prime}} 
		\Psi^{}_{\vec{n}^\prime \otimes \vec{n}^{\prime\prime}} \right)
		\ket{\vec{n}^{}}_A \bra{\vec{n}^\prime}
	\ ,
\end{equation}
where the occupation vectors $\vec{n} \equiv \{ n_m\}_{m \in A, \bar{A}}$
describe the configurations of particle numbers in the two subsets of DLL orbitals and the coefficient $\Psi_{\vec{n} \otimes \vec{n}^{\prime\prime}}$ is the amplitude of the component $\ket{\vec{n}}_A \otimes \ket{\vec{n}^{\prime\prime}}_{\bar{A}}$ within the multiparticle state $\Psi$.
The obtained density matrix is block-diagonal with respect to
the number of particles $N_A = \sum_{m \in A} n_m$ caught inside the orbital subset
and their angular index $L_A = \sum_{m \in A} m \, n_m$.
When the $M_A$ orbitals are the lowest lying ones,
this almost corresponds to a partition in real space:
the spectrum of $\rho^{(O)}$ thus carries precious information
on the edge excitations, without need of recurring to a detailed analysis of the energy spectrum.

The particle partition instead selects a subset of $N_A$ particles, 
regardless of their position in real or orbital space~\cite{bernevig11}.
It should therefore carry information on the bulk and excitation properties of the FQH state.
The density matrix reads:
\begin{equation} \label{eq:PES}
	\rho_{A}^{(P)} = \left( \bra{\psi} \prod_{j \le N_A} a^\dagger_{m_j} \, \prod_{k \le N_A} a^{}_{m_k^\prime} \ket{\psi} \right)
	\ket{\vec{m}} \bra{\vec{m}^\prime}
	\ ,
\end{equation}
where the configuration vectors $\vec{m} = \{m_1, \ldots, m_{N_A} \}$ (resp. $\vec{m}^\prime$)
scan over the orbitals of the lowest DLL, with the conserved quantity $L_A = \sum_{j \le N_A} m_j$
(and the convention $m_j > m_{j+1}$).

\begin{figure*}[!ht]
\includegraphics[width=0.475\textwidth]{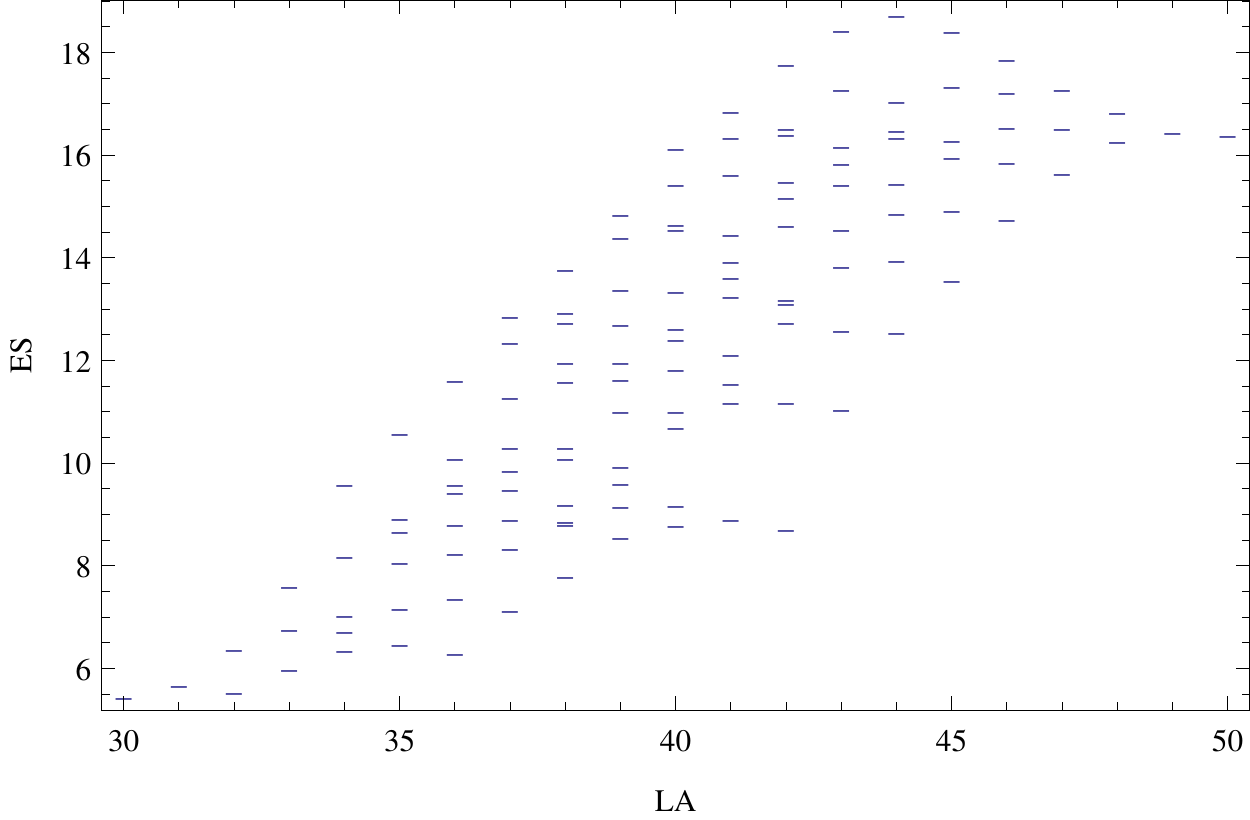}
\includegraphics[width=0.475\textwidth]{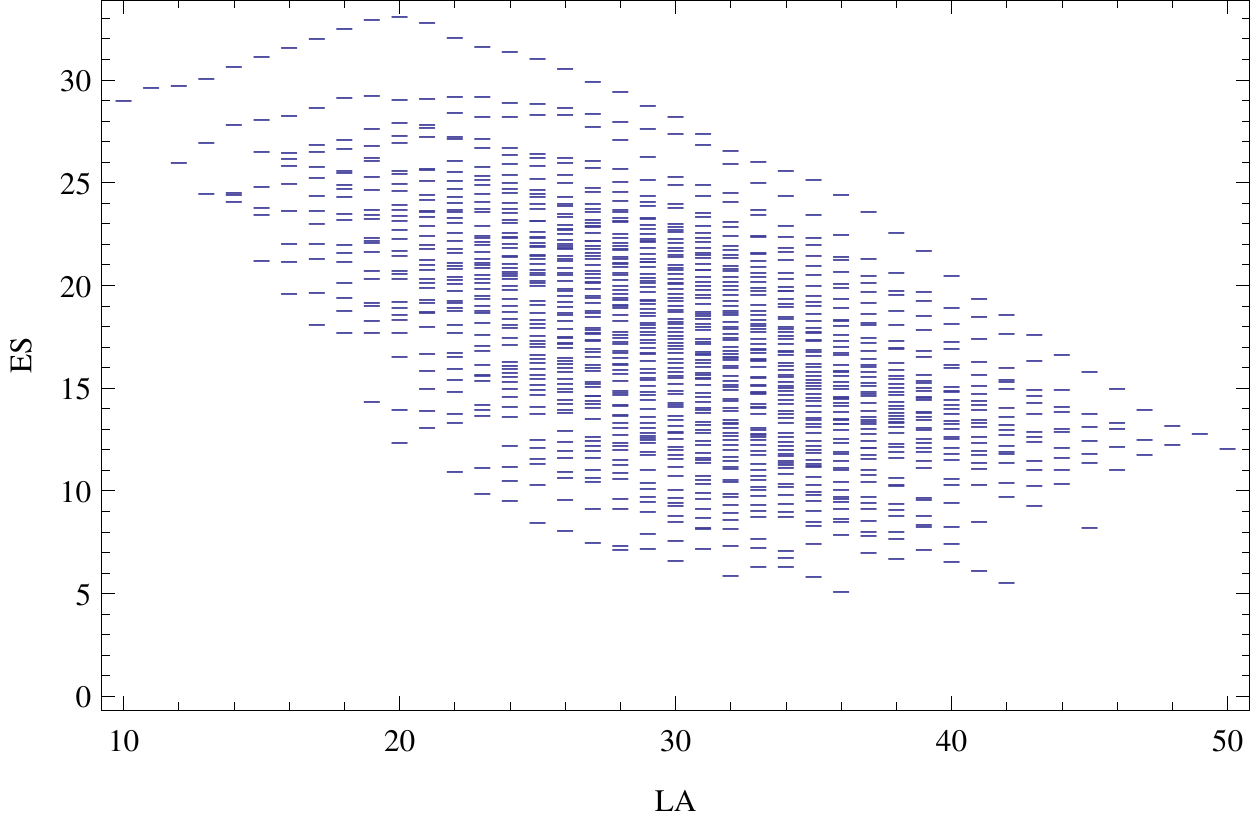}
\caption{Comparison of the orbital entanglement spectrum for the ground states at $L_{\rm tot}=135$ and $N=10$ for $q^2=0.5B$ (left) and $q^2=3.8B$ (right). The panel on the left shows the usual behaviour of a Laughlin state at filling $1/3$. The corresponding state in the second DLL (right) appears to be radically different and, in particular, the decreasing behaviour of the entanglement spectrum as a function of $L_A$ suggest an inversion of the chirality of the edge modes. The orbital entanglement spectrum is obtained by removing $m_A=12$ orbitals and $N_A=5$ atoms and it is plotted as a function of the residual angular momentum $L_A$.}
\label{fig:ESLaughlin}
\end{figure*}

\begin{figure*}[!ht]
\includegraphics[width=0.475\textwidth]{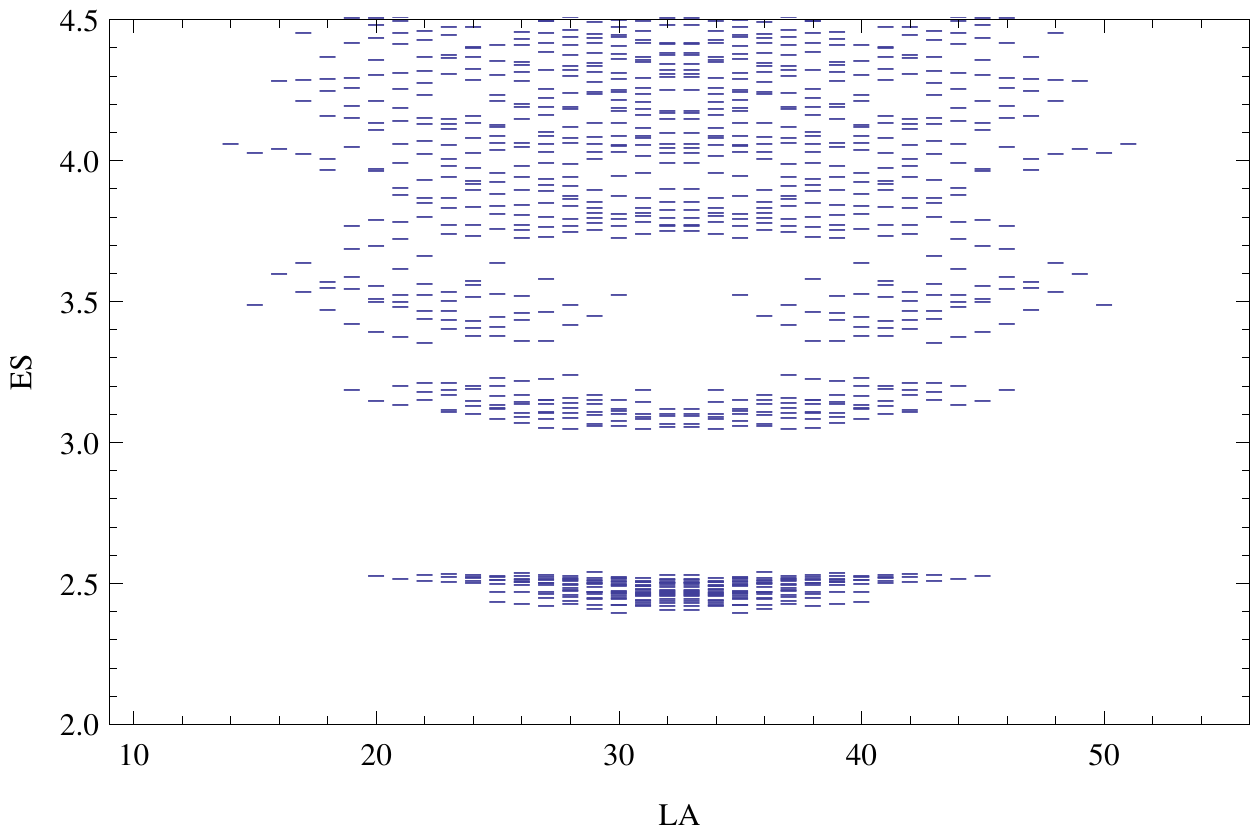}
\includegraphics[width=0.475\textwidth]{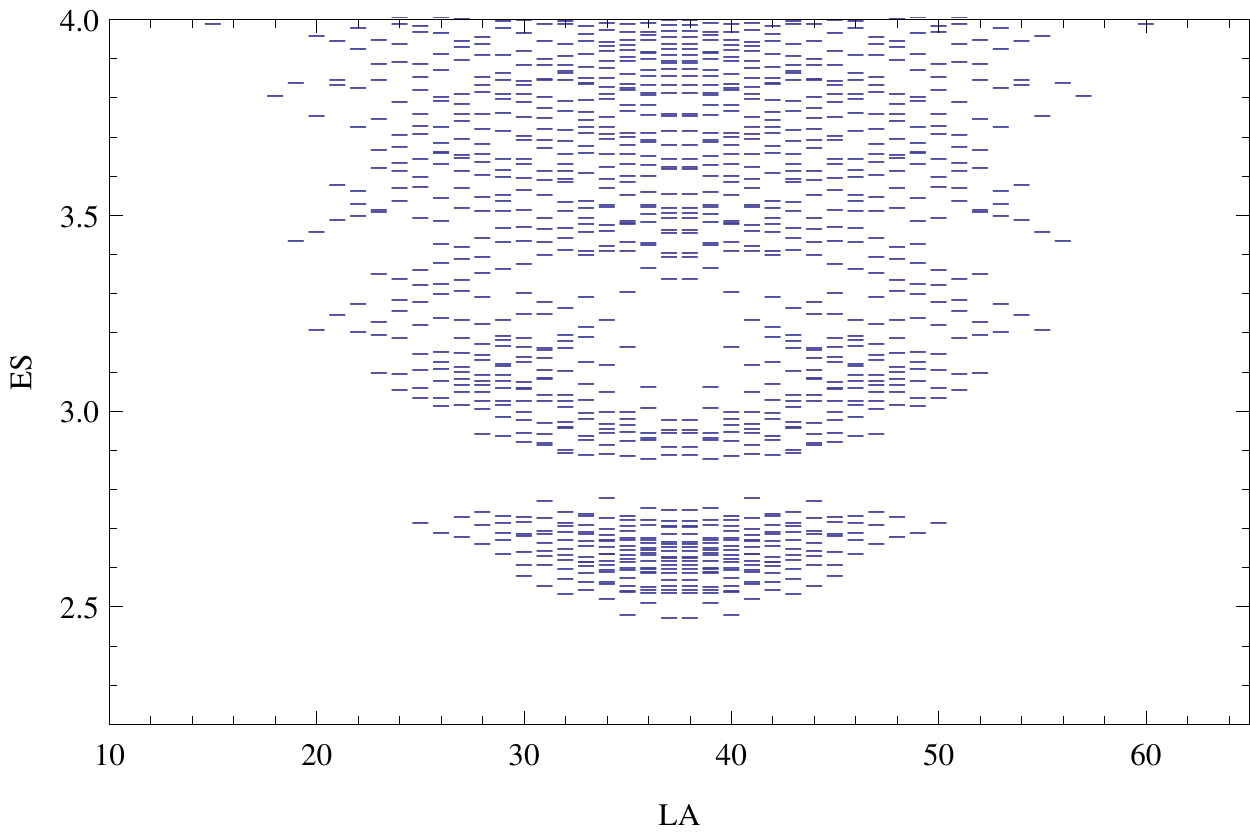}
\caption{\label{fig:esvortex} Particle entanglement spectrum of the states with 10 atoms at $q^2=3.8$, with $N_A=5$ at $L_{\rm tot}=65$ (left) and $L_{\rm tot}=75$ (right). The states are characterized by the presence of 2 and 3 vortices respectively. A gap separating the universal and  non-universal parts of the spectrum is evident. We notice, in particular, that the universal part appears at $L_{A,{\rm min}}=N_A(N_A-1)/2+2N_A$ and $L_{A,{\rm min}}=N_A(N_A-1)/2+3N_A$ respectively, consistently with the wavefunction for multiple vortices.} 
\end{figure*}

\begin{figure*}[!ht]
\includegraphics[width=0.475\textwidth]{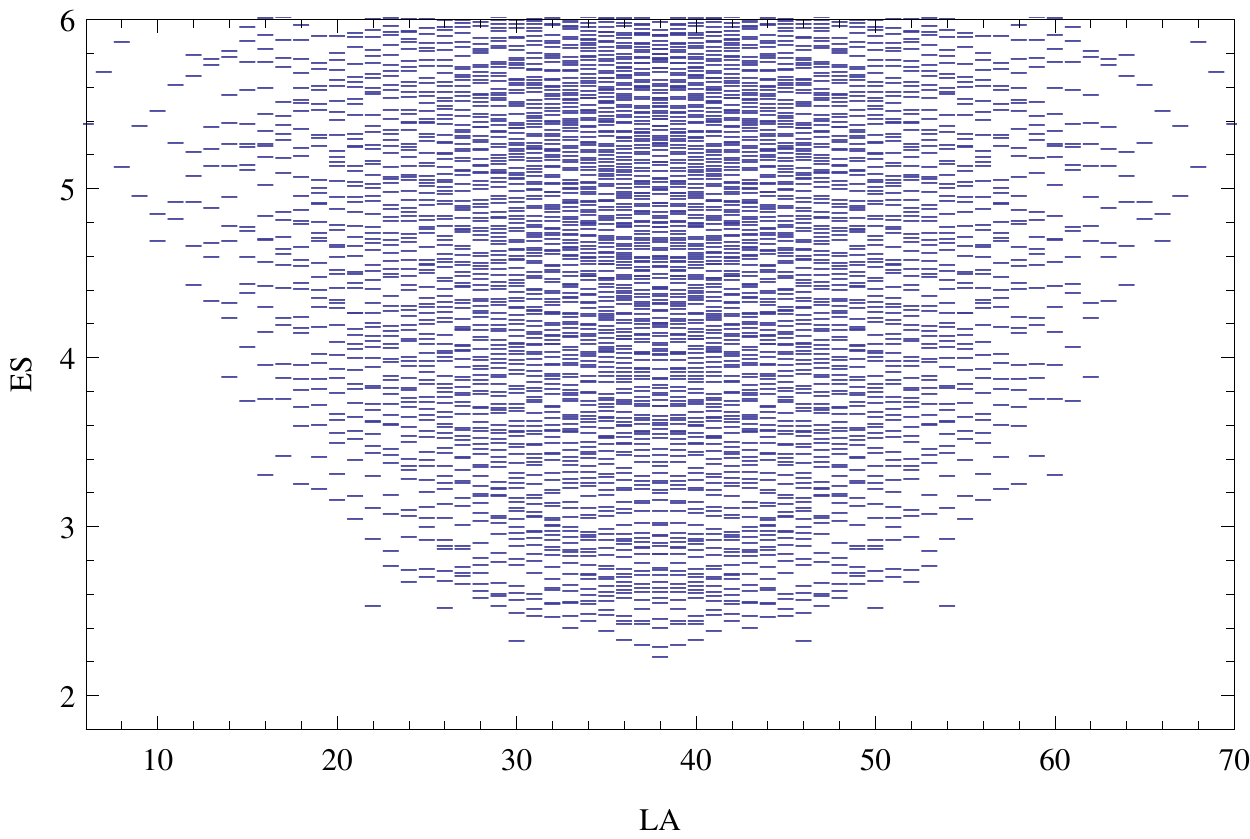}
\includegraphics[width=0.475\textwidth]{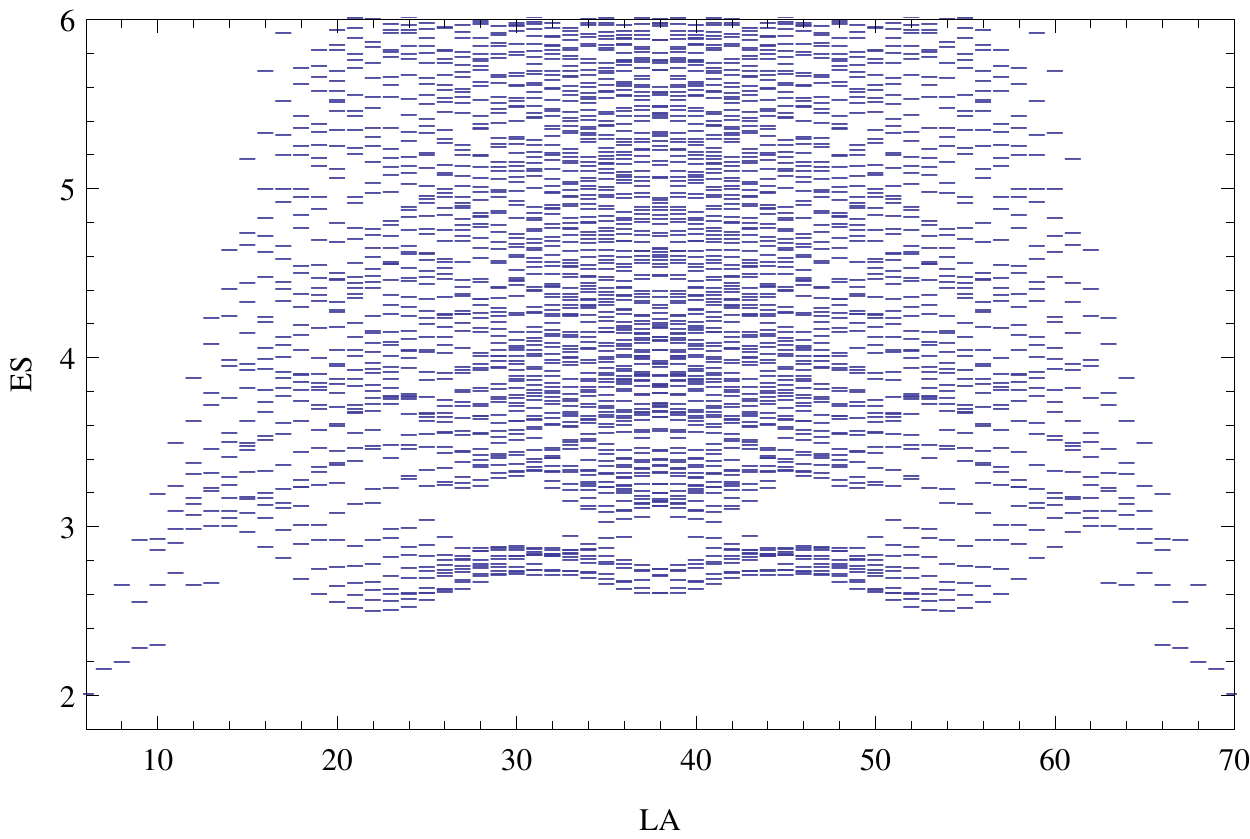}
\caption{Particle entanglement spectrum of the ground state with $N=8$ at $L_{\rm tot}=76$ and $q^2=3.8B$. Left panel: spectrum with $N_A=4$. Right panel: spectrum obtained in the conformal limit \cite{thomale10} under the same conditions.}
\label{fig:eshaffnian}
\end{figure*}

Our results of the previous Section~\ref{above} point to the existence 
in the second DLL of a new class of strongly correlated states, arising at intermediate density between
the $1/5$ Laughlin state (more dilute) and the vortex states (denser).
Since the full characterization of such states is a challenging task,
we decided to perform a systematic computation of orbital and particle ES 
to possibly shed new light on their nature.
We also extended such investigation to all the correlated states discussed above
in the first (Sec.~\ref{below}) and second DLL (Sec.~\ref{above}).
Our findings are summarized in Figs. \ref{fig:ESLaughlin}-\ref{fig:eshaffnian}:
The orbital ES gives, as expected, an extremely clear characterization of the
$\nu=1/3$ (in the first DLL) and $\nu=1/5$ (in the second DLL) Laughlin states.
The particle ES is better suited for the description of the vortex states appearing in the second DLL at high density;
for the strongly correlated paired states, the orbital ES turns out to be practically useless 
and we resorted to the computation of their particle ES in the conformal limit~\cite{thomale10}, i.e. by using the transformed wavefunction:
\begin{equation}\label{eq:conformalPES}
	\psi^{\rm (CL)}_{\vec{m}} \equiv  \psi_{\vec{m}} \, \cdot \prod_{j \le N} \sqrt{m_j !} \, ,
\end{equation}
apart from overall normalization factors. Such transformation is aimed at reducing the effect of the geometry of the system and emphasize the universal features of the states under investigation.

For the Haffnian-like states, this is the first time (to the best of our knowledge) that such 
an approach has been tried.

In Fig. \ref{fig:ESLaughlin} we plot the orbital ES for the ground states in the first and second DLL corresponding to the total angular momentum of the Laughlin state at filling $1/3$. 
As mentioned in the previous section this Laughlin state disappears in the second DLL as testified by the difference between the two spectra. 
In particular in the first DLL the orbital ES of the Laughlin state is exactly reproduced and the counting of the levels is related to the well-known counting of the degeneracies of its edge modes \cite{wen}, $1,1,2,3,5,7,11,\ldots$, even though corrections due the finite size effect are present \cite{hermanns11}. 
In the second DLL the behaviour of the ES appears to be inverted, opening the possibility of edge modes propagating with the opposite chirality. Interestingly this has been observed also in non-interacting lattice models with the same gauge potential \cite{burrello13}, where the non-Abelian term drives the system through a topological phase transition changing the chirality of the edge states.

Fig. \ref{fig:esvortex} provides a very clear characterization of the vortex states in the second DLL obtained by computing the particle ES. 
A marked gap between the lower universal part of the spectrum and the non-universal contribution on the top appears in a very visible way.
The universal part is consistent with a trial wavefunction proportional to $\prod_i z_i^m \prod_{i<j} (z_i-z_j)$ for the $m^{\rm th}$ vortex as may be observed by the minimum and maximum angular momentum of this part of the spectrum.
In particular $L_{A,{\rm min}}=mN_A+N_A(N_A-1)/2$ and $L_{A,{\rm max}}=L_{A,{\rm min}}+N_A(N-N_A)$.

Finally we investigated the ES for the states in the intermediate regime in the second DLL.
The results for the orbital ES are difficult to interpret and they do not present any universal feature or gap.
We rather considered the particle ES for $N=8$ and $L_{\rm tot}=76$, compatible with the Haffnian state.
Again the particle entanglement spectrum for $N=8$ and $N_A=4$ does not show a gap or other particular elements apparently ascribable to the wavefunction \eqref{haff}. In particular, by evaluating the minimum total angular momentum $L_{A,{\rm min}}$ of a subset of $N_A$ atoms in Eq. \eqref{haff}, one could expect a universal gap opening at $L_{A,{\rm min}}= 3N_A(N_A-1)/2-N_A$. This is barely seen in our numerical outcome possibly due to effect of the small number of particles to which the particle ES is rather affected \cite{ivan12}.
To reduce the finite size effects, however, we applied the conformal limit \cite{thomale10} to the particle ES: 
in this way one may distinguish an emergent gap for intermediate values of $L_A$ and a generic fermionic behaviour close to the minimal and maximal values of $L_A$.
Even though the conformal limit makes more clear the obtained results, we conclude that further investigations are needed for a full characterization of this Haffnian-like state.

\section{Conclusions}\label{concl}

In this work, we studied an interacting two-component fermionic gas,
subjected to an artificial $U(2)$ non-Abelian potential,
consisting of both a spin-orbit coupling and a magnetic field. 
This gauge potential preserves a Landau level (LL) structure,
though altering its form and causing crossings between different deformed 
(and spin-mixed) LL (DLL).
We investigated the system in the lowest DLL approximation,
far from the crossing points of the single-particle spectrum.
Extensions closer to such degeneracy points are feasible~\cite{grass12a,grass12b}
and would require more refined techniques to account for
 LL mixing~\cite{simon12,nayak09,nayak13,simon13,macdonald13,quinn13,nayak14}.

We discussed the role of the trapping potential and showed that,
when combined with a Zeeman term, it can give rise to an angular momentum term,
which can in turn be used to test the stability of the obtained correlated states.
Further corrections to the DLL structure and to the effective interactions
are discussed as being almost negligible.

Once projected onto the DLL, the $s$-wave interspecies contact interactions translate into
an effective short-range interaction, described by a finite number of Haldane pseudopotentials (HP).
These HPs display a peculiar growth with the relative angular momentum,
and are therefore dubbed as {\it non-monotonic} (NMHP, see Fig.~\ref{tab:Wn}).
Their number depends on the character of the lowest DLL, 
i.e., on the intensity $q$ of the spin-orbit coupling with respect to the above mentioned DLL crossing points.
For small $q$'s the system is mapped onto a hard-core model, 
correctly described by a composite fermion picture:
the Laughlin $\nu = 1/3$ state appears as the global ground state 
for weak trapping potential, followed by the formation of 
Jain hierarchical states once the confinement is increased.
This shows that the interplay between contact interaction and the non-Abelian gauge potential is indeed sufficient to obtain spin-polarized interacting ultracold fermions, thus enabling the formation of a Laughlin state without long-range interactions, as dipolar ones.
For $q$'s beyond the first crossing point, instead, the HPs atypical inversion
($W_3 / W_1 = 3$ in the second DLL analysed here)
considerably changes the picture with respect to ordinary 
fractional quantum Hall systems in solid state setups.
We stress that, for fermions, no true long-range dipolar interaction is needed 
to attain such inversion, as it would instead be in bosonic mixtures~\cite{grass13}, thus making the ultracold fermionic gases in non-Abelian potentials a very promising playground to obtain new classes of strongly correlated states. 

The large $W_3 / W_1$ ratio is responsible for three qualitatively different phases
in the second DLL, once we increase the trapping parameter, and thus the density.
We examined their distinctive features in experimentally observable quantities,
such as the (spin-resolved) density profiles and correlation functions.
First, at low trapping the $\nu = 1/3$ Laughlin states disappears from the stabilized ones,
in favour of the even more dilute $\nu = 1/5$ Laughlin states, usually elusive in solid-state realizations.

Then, at intermediate densities we obtain different strongly-correlated states,
characterized by an effective pairing among the atoms due to the HP inverted ordering:
their two-body correlation functions are reminiscent of the Haffnian $\nu = 1/3$ trial states.

Around the maximally dense integer quantum Hall state,
the system is characterized by the presence of vortices,
whose character qualitatively changes at a critical number of particles $N_c$,
which we identify to be $N_c=13$.
Below that threshold, relevant for microtrapping, states with central 
vortices are favoured and the non-Abelian potential even gives rise to skyrmionic spin textures.
For larger population the introduction of central vortices becomes unfavourable 
and the system may be driven to different vortex lattice geometries,
however not precisely identifiable in our axially symmetric ansatz. 

In order to capture the nature of these three classes of states,
we resorted as well to quantum information tools,
namely we studied the orbital and particle entanglement spectrum (ES).
The ES gave a clear characterization of Laughlin states (orbital ES)
and vortex states  (particle ES),
while information on Haffnian-like states can be obtained 
by performing the  conformal limit of the particle ES:
however, more extensive numerical investigations are required 
to have a satisfactory characterization 
of the intermediate incompressible states in the second DLL.

Another possible tool of verifying the nature of such states would be to calculate their overlap with the corresponding analytical expressions. 
This could be achieved either from the evaluation of multi-dimensional integrals in real space, based on the second-quantized wavefunctions or by translating these wavefunctions into a suitable expression in the Fock basis. An alternative strategy could be also to consider second-quantized parent Hamiltonian for the target wavefunctions such as the Haffnian state. We leave this analysis for future works.

To conclude, let us illustrate some possible extensions of the present work,
aiming to proceed further in the creation, observation and possibly control
of atypical fractional quantum Hall states and their excitations,
thanks to the flexibility of ultracold atomic platforms.
A first direction would be 
the study of correlated states in the next deformed levels and 
the analysis of the ground-states at the degeneracy points.
Another direction would be to consider a simpler setup 
where the spatially varying Zeeman term is absent in \eqref{ham2_in} or not exactly compensated as in \eqref{additional}:
an extra term added to the angular momentum term in \eqref{ham2} would be present, which 
could be exploited as a new ingredient for the state engineering. Furthermore, additional Zeeman terms would change the single-particle wavefunction and consequently affect the effective interactions, as verified in lattice systems with synthetic magnetic fields \cite{maska,cicci}. Our analysis could also be extended and compared with results obtained for different geometries or trapping potentials. In particular it could be interesting to evaluate the behavior of the ground states in the presence of hard-wall boundary conditions such the ones experimentally obtained for ultracold bosons in \cite{gaunt13,corman14,corman14b}.
A steeper but potentially richer direction could be the investigation of general
(LL preserving) non-Abelian gauge potentials, like the ones with general Rashba and Dresselhaus terms which can be mapped into the single particle models studied in \cite{gritsev,amico}.

\textit{Acknowledgements} - We warmly thank Xin Wan, Jiannis Pachos, Tobias Grass, Ivan Rodriguez and Ying-Hai Wu
for useful discussions. M. Burrello acknowledges support from the EU grant SIQS,
M. Rizzi from a Stufe-I project by the JGU, 
M. Roncaglia from the Compagnia di S. Paolo and A. Trombettoni from the FET MatterWave.
The numerical simulations were run on the MOGON cluster of JGU-Mainz.

\appendix
\section{Decomposition of two-body Fock states into  eigenstates of the relative angular momentum}\label{app_dec}

The formulation of the second-quantized Hamiltonian~\eqref{twobody}
requires the determination of the coefficients $g[m, M, j]$.
They describe a basis transformation from the Fock states of two fermions
with total momentum $M$, i.e. $\ket{j, \, M-j}$,
into the two-body wavefunctions with relative angular momentum $m$,
which are then eigenstates of the pseudopotential interaction.
Since the deformation of the LL caused by the non-Abelian potential
is already taken into account in the HP definition~\eqref{eqwn},
the required coefficients $g$ must be calculated starting from the normalized
two-body wavefunction in a {\it canonical} lowest Landau level:
\begin{equation}
\psi_{j,M-j}\left(z_{1},z_{2}\right) = z_{1}^{j}z_{2}^{M-j}
\frac{e^{-\frac{1}{4}\left(\left|z_{1}\right|^{2}+\left|z_{2}\right|^{2}\right)}}{2^{M/2}\, 2\pi\sqrt{j!\left(M-j\right)!}}
\end{equation}
where the coordinates are expressed in units of the magnetic length.

As a first step we decompose the wavefunction in terms of the center of mass and relative coordinates:
\begin{equation}
Z=\frac{z_{1}+z_{2}}{2}\,,\qquad z=z_{1}-z_{2} \, .
\end{equation}
The coordinate monomials in $\psi$ can then be rewritten as: 
\begin{equation}
z_{1}^{j}z_{2}^{M-j}=\sum_{p=0}^{j}\sum_{q=0}^{M-j}Z^{M-q-p}\frac{z^{p+q}}{2^{p+q}}(-1)^{q}
\begin{pmatrix} j\\ p \end{pmatrix}
\begin{pmatrix} M-j\\ q \end{pmatrix}
\,,
\end{equation}
where the relative angular momentum $m = p+q$ is now explicitly visible.
The double sum can be expressed as 
\begin{equation*}
	\sum_{p=0}^{j}\sum_{q=0}^{M-j} \to 
	\sum_{m=0}^{M} \ \sum_{p=\max(0,m-M+j)}^{\min\left(j,m\right)} \, .
\end{equation*}
We notice also that the Gaussian exponent reads 
$(\left|z_{1}\right|^{2}+\left|z_{2}\right|^{2}) /4 = \left|Z\right|^{2}/2 + \left|z\right|^{2}/8$,
and the normalisation coefficients of the new monomials thus become
$\sqrt{\pi \, (M-m)!}$ for $Z^{M-m}$ and $2^m \sqrt{4 \pi \, m!}$ for $z^{m}$, respectively.

When we put all these rewritings together, we obtain the desired transformation:
\begin{equation}
\small
\psi_{j,M-j} \to \sum_{m=0}^{M} \,
			g[m,M,j] \, \left[  \frac{Z^{M-m} e^{-|Z|^2/2}}{\sqrt{\pi (M-m)!}} \right] \,
			\left[  \frac{z^{m}e^{-|z|^2/8}}{2^m \sqrt{4 \pi m!}} \right]
\end{equation}
with:
\begin{equation}
\small
g\left[m,M,j\right]
=\hspace{-0.5cm}\sum_{p=\max(0,m-M+j)}^{\min\left(j,m\right)}
\sqrt{\frac{\left(M-m\right)!m!}{j!\left(M-j\right)!}}\frac{(-1)^{m-p}}{2^{M/2}}
\begin{pmatrix} j\\ p \end{pmatrix}
\begin{pmatrix}M-j\\ m-p \end{pmatrix}
\end{equation}
where the following normalization holds: 
\begin{equation}
\sum_{m=0}^{m=M}\left(g[m,M,j]\right)^{2}=1 \, .
\end{equation}
In particular: 
\begin{align*}
g[1,M,j] & =  -\frac{(M-2j)}{2^{M/2}}\sqrt{\begin{pmatrix}M\\
j
\end{pmatrix}M^{-1}}\\
g[3,M,j] & =  \frac{\left((M-2j)^{2}-3M+2\right)\left(2j-M\right)}{2^{M/2}\cdot6}\sqrt{\frac{\begin{pmatrix}M\\
j
\end{pmatrix}}{\begin{pmatrix}M\\
3
\end{pmatrix}}}
\end{align*}

\section{Numerical procedure}\label{app_num}

In order to conveniently treat the (immense) 
Hilbert space of the many-body system,
we exploit the block-diagonal form of the Hamiltonian~\eqref{twobody}
and the fermionic nature of the particles. 
To generate the yrast spectrum is then sufficient to consider {\it only}
the possible configurations of $N$ distinct integers (i.e. the single particle momenta $m_j$, $j=1 \ldots N$)
to sum up to a given total $L$ (the effective total angular momentum).
Moreover, from physical considerations, we can also cut the available values of such integers
at an upper threshold $\bar{m} \simeq N / \nu$, where $\nu \simeq O(N^2 / 2L)$ is the filling factor.
These ``integer partitions'' can be generated efficiently in an iterative way;
their number grows roughly as
$5.8 \times 10^{0.69 N - 2}$ for the Laughlin $\nu=1/3$ momentum sector ($\simeq 6 \cdot 10^5$ for $N = 10$)
and as $2.7 \times 10^{0.94 N - 2}$ for the Laughlin $\nu=1/5$ one ($\simeq 10^6$ for $N = 8$).

We must account for the fermionic character of the particles also
when writing the matrix representation of the Hamiltonian over this basis.
To this aim, it is more convenient to go back to the Fock representation
and perform a Jordan-Wigner mapping with the following convention:
\begin{equation}
	\ket{n_0, \ldots, n_{\bar{m}}} =
	(a^\dagger_{\bar{m}})^{n_{\bar{m}}} \ldots (a^\dagger_{\bar{0}})^{n_{\bar{0}}}
	\, \ket{0, 0, \ldots, 0} \, .
\end{equation}
The terms $a^\dagger_{M-m_1} a^\dagger_{m_1}$ (with $m_1 \le \lfloor M/2 \rfloor$)
therefore get a pre-factor sign depending  on the parity of  $\sum_{j=m_1+1}^{M-m_1-1} n_j$.

An even more important point for computational purposes is the sparseness of the so obtained matrix,
containing only less than a hundred of elements per row, in the cases considered here.
To get a flavour of the computational resources needed,
the diagonalization of the $N=8$ Laughlin $1/5$ state at $L=140$ takes around $5$ hours and around $4$Gb of RAM.
We finally notice that the computation of reduced density matrices (and their diagonalization)
is per se a more modest task but, due to the combinatorial growth of instances, 
its total cost could even overcome the Hamiltonian solution itself.
For this reason, we presented the entanglement spectra for slightly smaller samples.

The Fortran code is available, upon request and mutual agreement,
for cross-checks and extensions of the present work.

\section{Transition to the integer quantum Hall states} \label{app_est}

In this Appendix we provide a rough estimate of the threshold of the parameter $\Delta/v$ separating the integer quantum Hall regime from the other stabilised states appearing in the yrast spectrum both in the first and second DLL.
To calculate this value we compare the total energy of the maximum density droplet, constituted by the sum of the interaction energy and the trap contribution, with the trapping energy $L_{\rm tot}\Delta$ of the Laughlin state with no interaction energy. This means that, for the first and the second DLL, we are considering the Laughlin states, respectively at filling $1/3$ and $1/5$, as good representatives of the fractional quantum Hall regimes for $q^2<q_1^2$ and $q_1^2<q^2<q^2_2$, thus neglecting the presence of the other intermediate states with positive interaction energy appearing in the system. This leads to an estimated critical value $\Omega_c$ that will underestimate, in general, the physical value of $\Omega$ at which the IQH droplet appears.

Let us consider first the case $q^2<q_1^2$. The $\nu=1/3$ Laughlin state has no interaction energy and a trapping energy $E_{1/3}= 3N(N-1)\Delta/2$. Considering the integer quantum Hall state, instead, there is a trapping energy $ N(N-1)\Delta/2$ and an interaction energy that can be approximated by:
\begin{multline} \label{estim}
 U = \frac{2}{\pi} \mathcal{G}\left( m_{\rm rel}=1\right) \frac{N(N-1)}{2} W^{(1)}_1 \approx \\
 \approx \frac{2}{\pi}(N-1) v\frac{\sin^2\varphi_1}{8} \approx \frac{(N-1)v}{5\pi}
\end{multline}
where $\mathcal{G}\left( m_{\rm rel}\right)$, called the pair amplitude, denotes the fraction of pairs of atoms of relative angular momentum $m_{\rm rel}$ within the IQH state (see, for example \cite{wojs04}). In particular, we roughly approximate the pair amplitude with a constant distribution in $m_{\rm rel}$, whereas the exact $\mathcal{G}$ should be evaluated starting from the factors $g\left(m_{\rm rel},M,j\right)$ and the correlation functions; therefore, we substitute $\mathcal{G}\left( 1\right)\approx 2/N$ with the effect of overestimating the interaction energy, as can be seen by comparing Eq. \eqref{estim} with the numerical results (the factor ${2}/{\pi}$ has been added for consistency with the energy scale adopted in the numerical data). In the second approximation we substituted the pseudopotential $W^{(1)}_1$ with its approximated value $v/10$ for $q^2=0.5B$, since $\Delta \ll 1$. 
Finally, by comparing the energies $U +  N(N-1)\Delta/2$ and $E_{1/3}$ we obtain, for $q^2=B/2$, the rough estimate:
\begin{equation} \label{deltac1}
 \frac{\Delta_c^{(1)}}{v}=\frac{1}{5\pi N}
\end{equation}
such that, for $N=10$, $\Delta_c \approx 0.0064v$. By looking at Fig. \ref{yrast10} we see that this value lies in one of the regular plateau preceding the appearance of the maximum density droplet. Such plateaus can be interpreted as excitations of the IQH state at filling $1$, therefore Eq. \eqref{deltac1} provides a rough estimate of the value of $\Delta$ bringing to the transition to the integer quantum Hall regime. We verified a similar behavior for $6\le N < 10$.

Let us consider now the behavior in the second DLL. Here the interaction becomes zero for the $\nu=1/5$ Laughlin state with trapping energy $E_{1/5}= 5N(N-1)\Delta/2$. Furthermore the estimate of the energy of the IQH droplet must account for both $W^{(2)}_1$ and $W^{(2)}_3$ thus leading, for $q^2=3.8B$, to:
\begin{multline} \label{estim2}
 U = \frac{2}{\pi} \left( \mathcal{G}\left(1\right)  W^{(2)}_1 + \mathcal{G}\left(2\right)  W^{(2)}_2 \right) \frac{N(N-1)}{2} \approx \\
 \approx \frac{2}{\pi}(N-1) v\frac{\sin^2\varphi_2}{8} \approx \frac{(N-1)v}{4\pi}
\end{multline}
where we adopted the previous assumption of equally distributed pairs, we exploited that $\sum_{m}W^{(n)}_m=\sin^2\left( \varphi_n\right) / 8$ and we considered $\sin^2 \varphi_2 \approx 1$ for $q^2=3.8B$. Also in this case, $U$ provides an overestimation of the interaction energy of the maximum density droplet that, however, improves with increasing $N$. From Eq. \eqref{estim2} we obtain:
\begin{equation} \label{deltac2}
 \frac{\Delta_c^{(2)}}{v}=\frac{1}{8\pi N}.
\end{equation}
For $N=8$ this results in $\Delta_c^{(2)} \approx 0.005 v$ which roughly predicts the critical value separating the intermediate FQH regime from the vertex regime discussed in Sec. \ref{above}. We verified that Eq. \eqref{deltac2} provides a reasonable value for the transition from the fractional to the vortex phase for all the numerical data at our disposal ($6\le N \le 10$).

Finally we observe that both the values $\Delta_c^{(1)}$ and $\Delta_c^{(2)}$ scale like $1/N$ thus implying the necessity of smaller and smaller confining potentials $\left(\Delta \approx \omega^2/(2B) \right)$  with increasing $N$ in order to obtain a fractional regime. This is consistent with having a constant energy gap $\mathcal{D}$ for the fractional quantum Hall states in the thermodynamic limit (see Fig. \ref{laughlin}).

\end{document}